\documentclass[useAMS,usenatbib]{mnras}
\usepackage{epsfig, bm, times, aas_macros}
\usepackage[english]{babel}
\usepackage{amsmath}
\usepackage{amssymb}
\usepackage{graphicx}
\usepackage{color}
\usepackage{threeparttable}

\newcommand{\seclab}{Section}
\newcommand{\tablab}{Table}

\newcommand{\figlab}{Figure}
\newcommand{\equlab}{Eq}
\newcommand{\secref}[1]{\seclab{} \ref{#1}}
\newcommand{\tabref}[1]{\tablab{} \ref{#1}}

\newcommand{\figref}[1]{\figlab{} \ref{#1}}

\newcommand{\eqr}[1]{\equlab{.} \eqref{#1}}
\newcommand{\eqrs}[1]{\equlab{s.} \eqref{#1}}

\newcommand{\tent}[1]{\times 10^{#1}}

\newcommand{\kep}{\texttt{KEPLER}}

\newcommand{\rmo}{m_{r\,1}}
\newcommand{\rno}{n_{r\,1}}
\newcommand{\rmt}{m_{r\,2}}
\newcommand{\rnt}{n_{r\,2}} 
\newcommand{\amo}{m_{\alpha\,1}}
\newcommand{\ano}{n_{\alpha\,1}}
\newcommand{\amt}{m_{\alpha\,2}}
\newcommand{\ant}{n_{\alpha\,2}}
\newcommand{\mc}{\log\dot{M}_{\rm{c}}}
\newcommand{\rc}{\log R_{\rm{c}}}
\newcommand{\ac}{\log \alpha_{\rm{c}}}
\newcommand{\lm}{\log \dot{M}}
\newcommand{\lr}{\log R}
\newcommand{\lal}{\log \alpha}
\newcommand{\fit}[3]{$#1^{+#2}_{-#3}$}
\newcommand{\etab}[2]{\begin{tabular}{@{}c@{}}#1\\(#2)\end{tabular}}
\newcommand{\nlbreak}{\\[0.95em]}
\newcommand{\minb}{MINBAR}
\newcommand{\sat}[1]{\textit{#1}}

\newcommand{\us}{4U 1636--536}
\newcommand{\ue}{4U 1608--522}
\newcommand{\aqu}{Aql X--1}
\newcommand{\exo}{EXO 0748--676}
\newcommand{\ks}{KS 1731--260}

\newcommand{\combi}[4]{
  #1\\
  $\phantom{a}$\hspace{\stretch{1}}
  \includegraphics[width=#2\textwidth]{#3.eps}
  \hspace{1cm}
  \includegraphics[width=#2\textwidth]{#4.eps}
  \hspace{\stretch{1}}$\phantom{a}$
}

\newcommand{\figres}{%
  \begin{figure*}
    \begin{center}
      \combi{\ks{,} $\nu = 524$}{ 0.28}{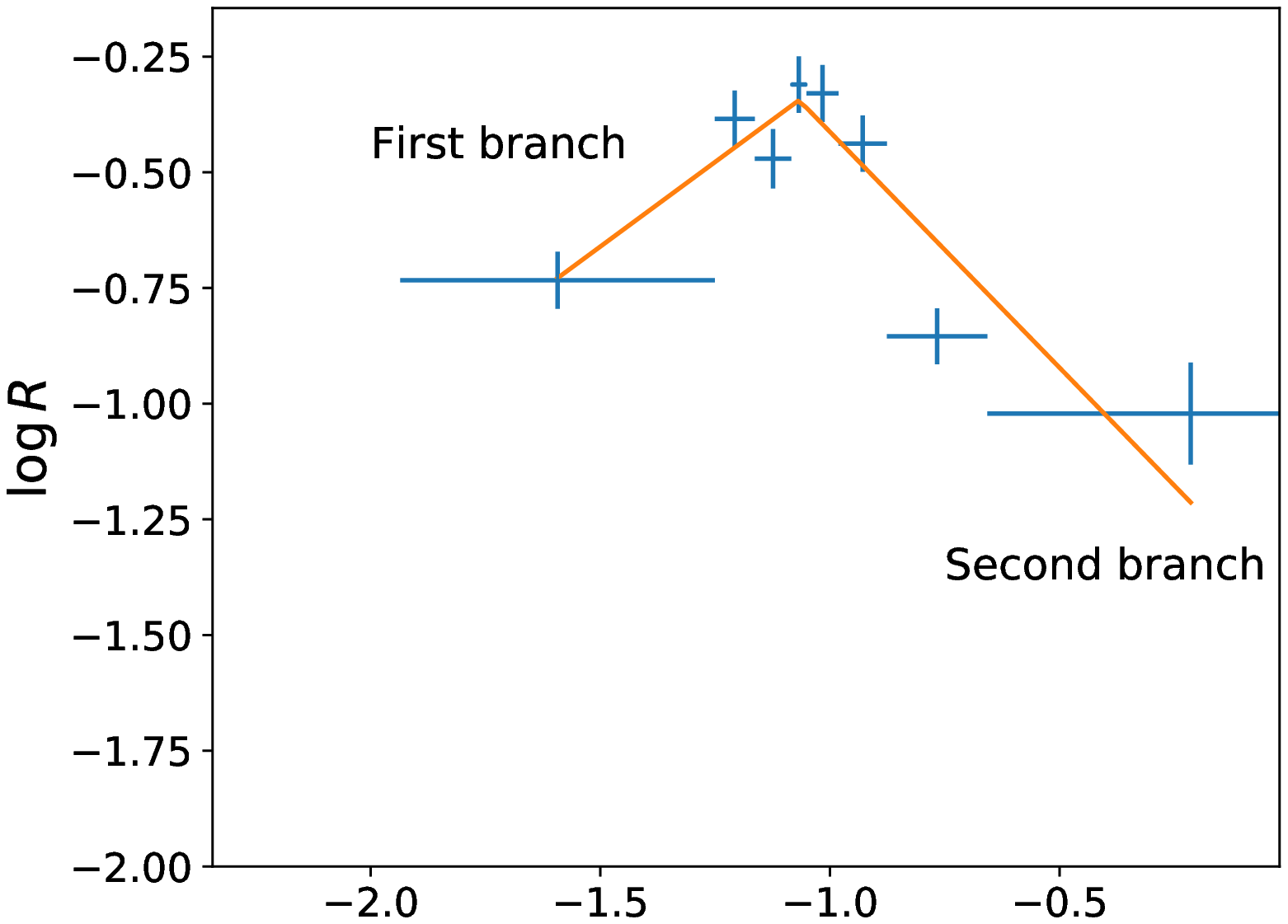}{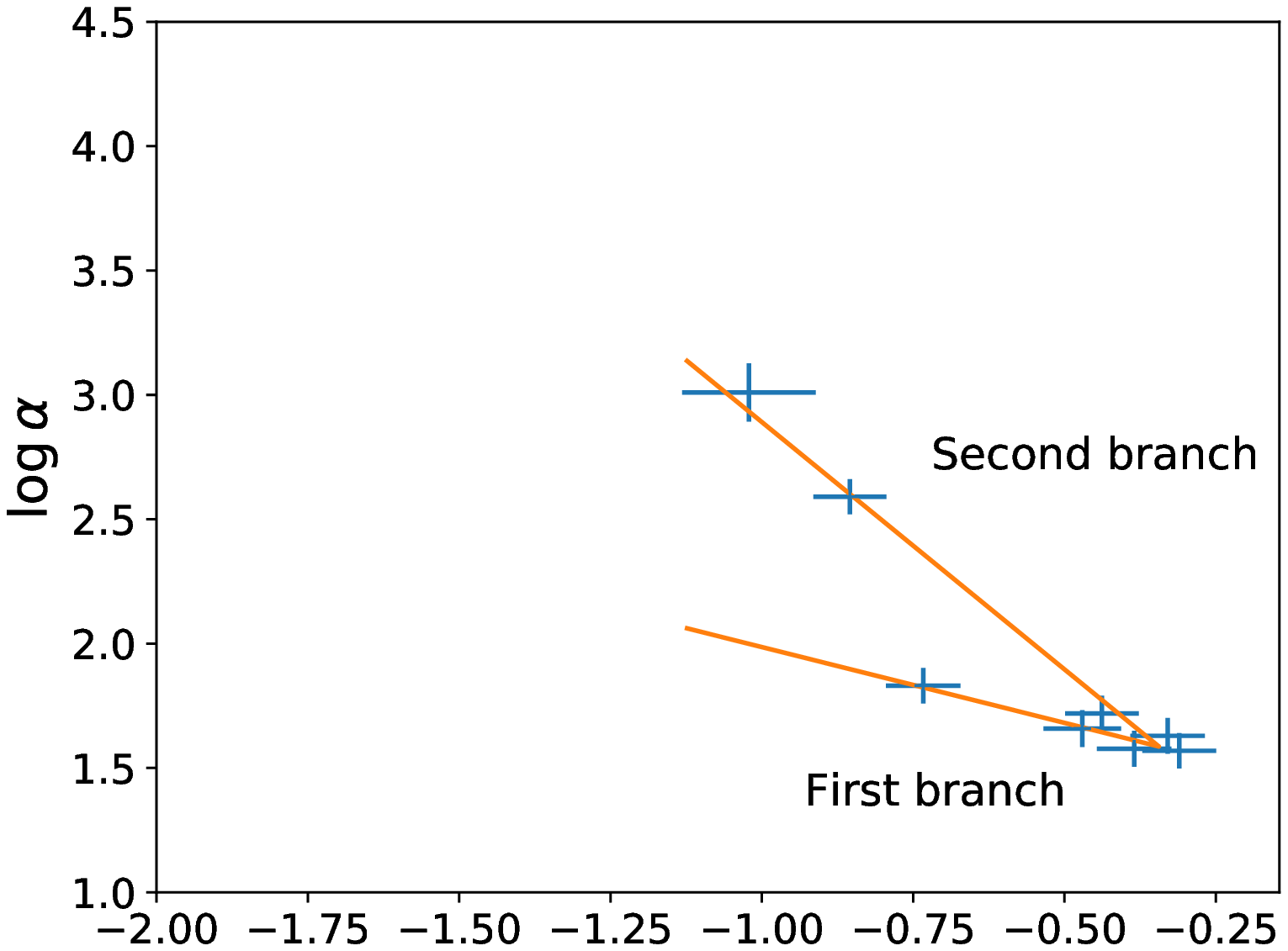}\\
      \combi{\aqu{,} $\nu = 549$}{0.28}{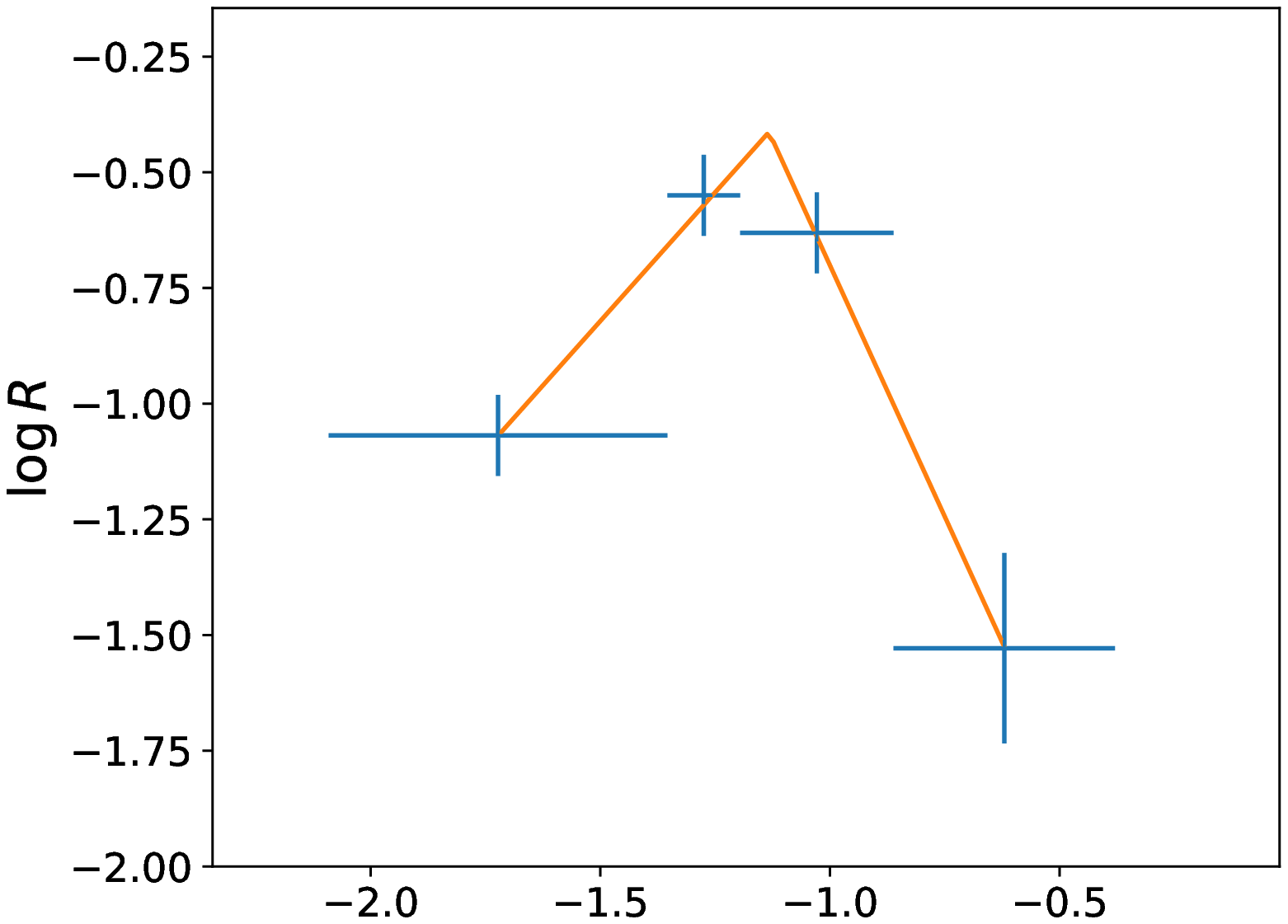}{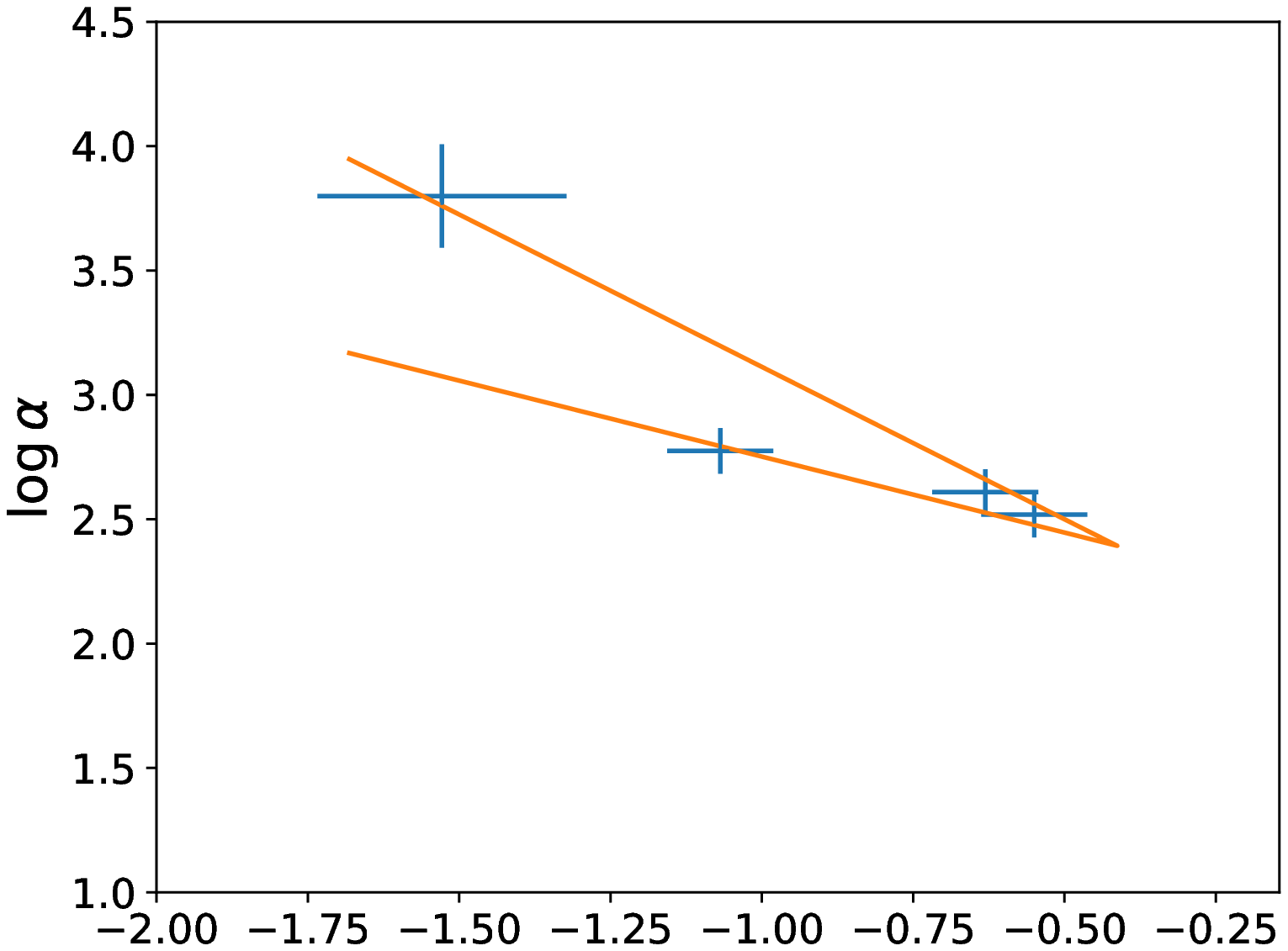}\\
      \combi{\exo{,} $\nu = 552$}{0.28}{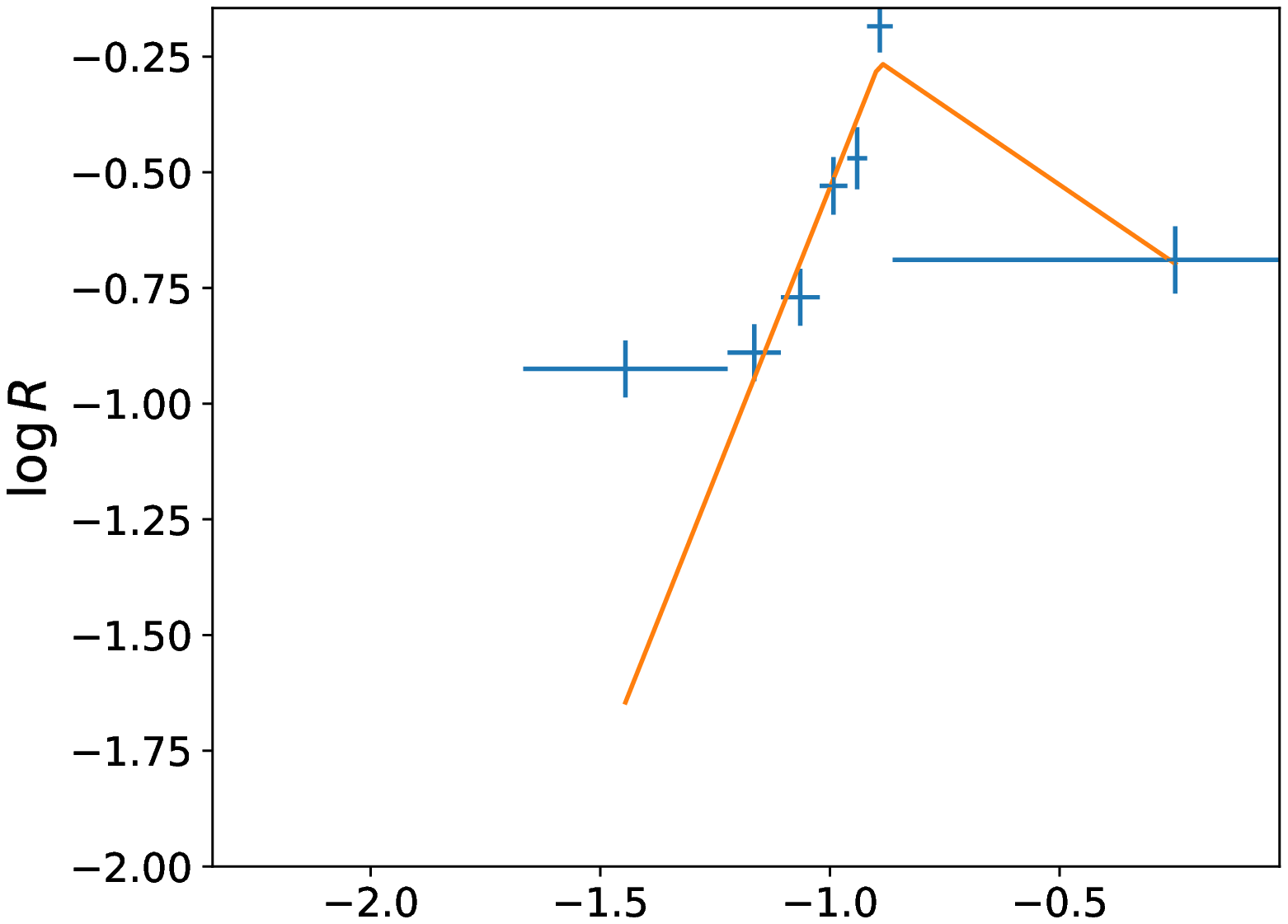}{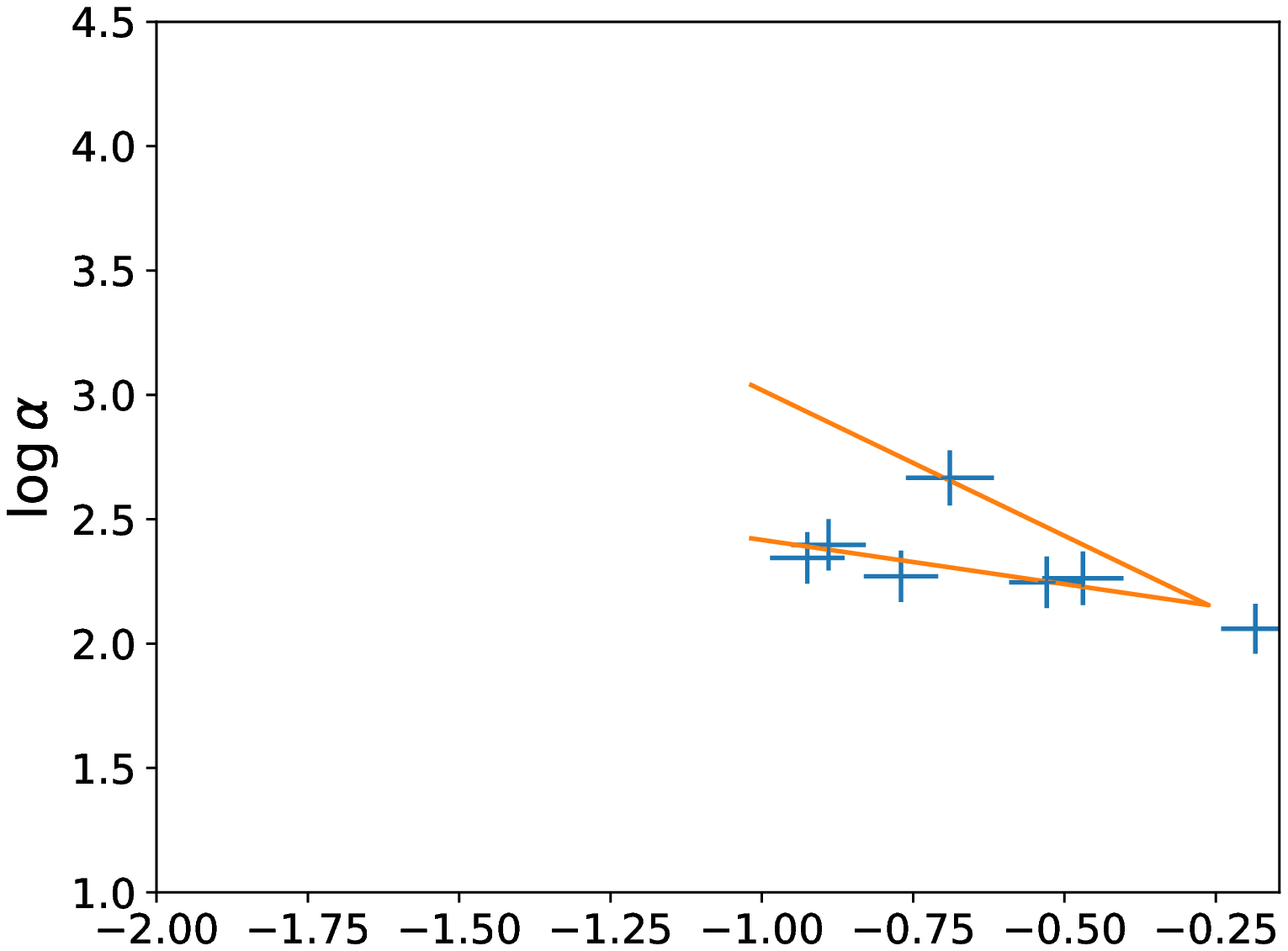}\\
      \combi{\us{,} $\nu = 581$}{ 0.28}{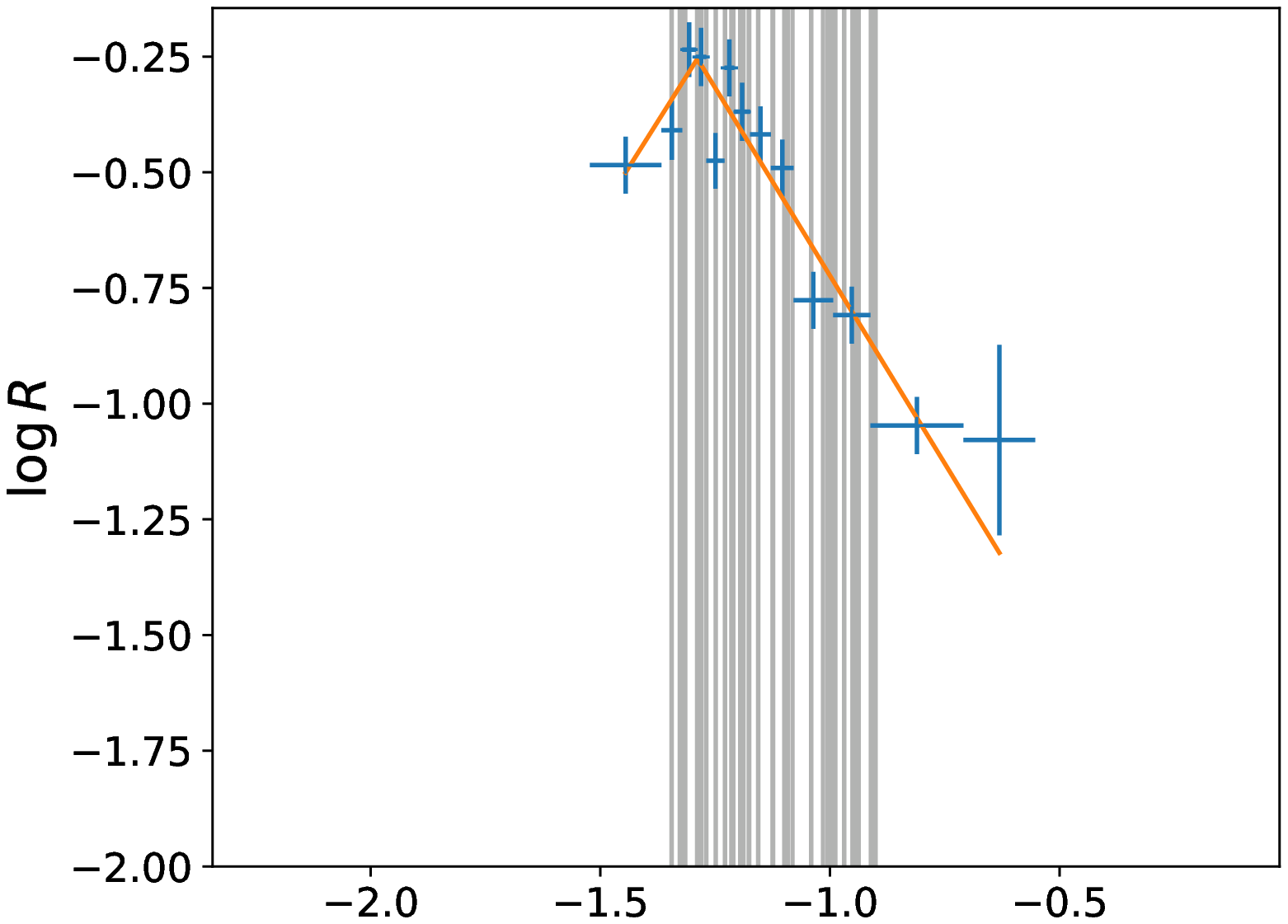}{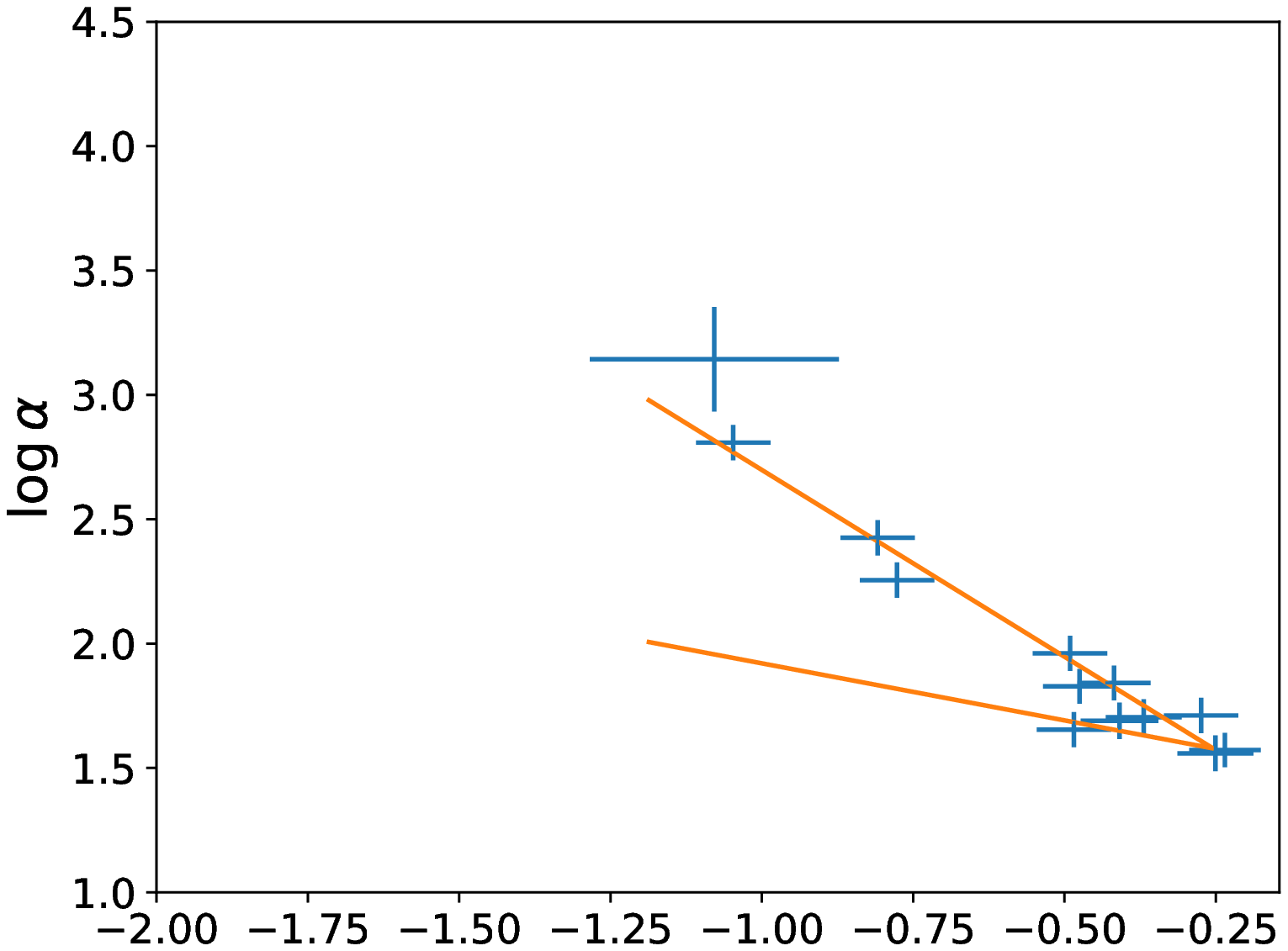}\\
      \combi{\ue{,} $\nu = 620$}{ 0.28}{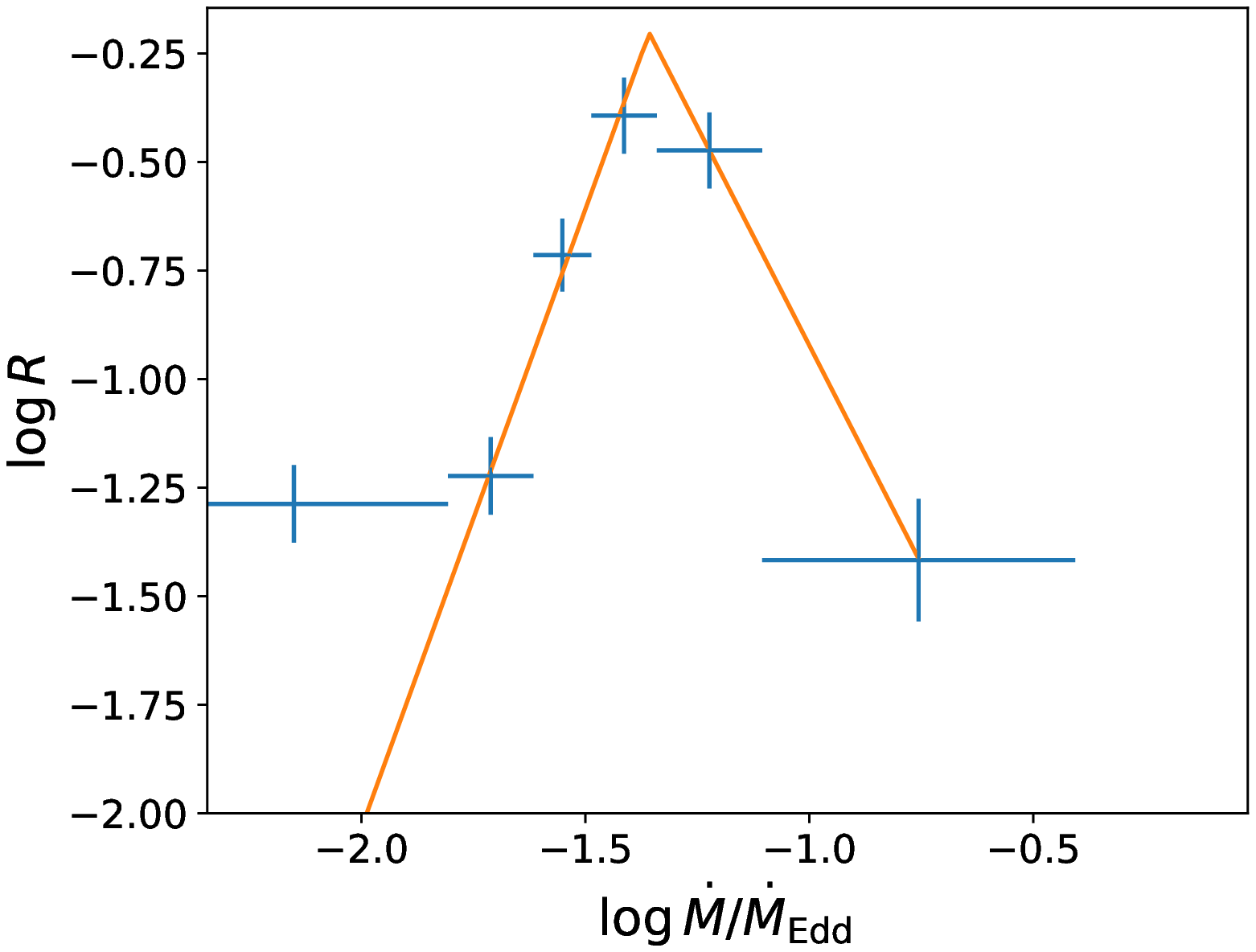}{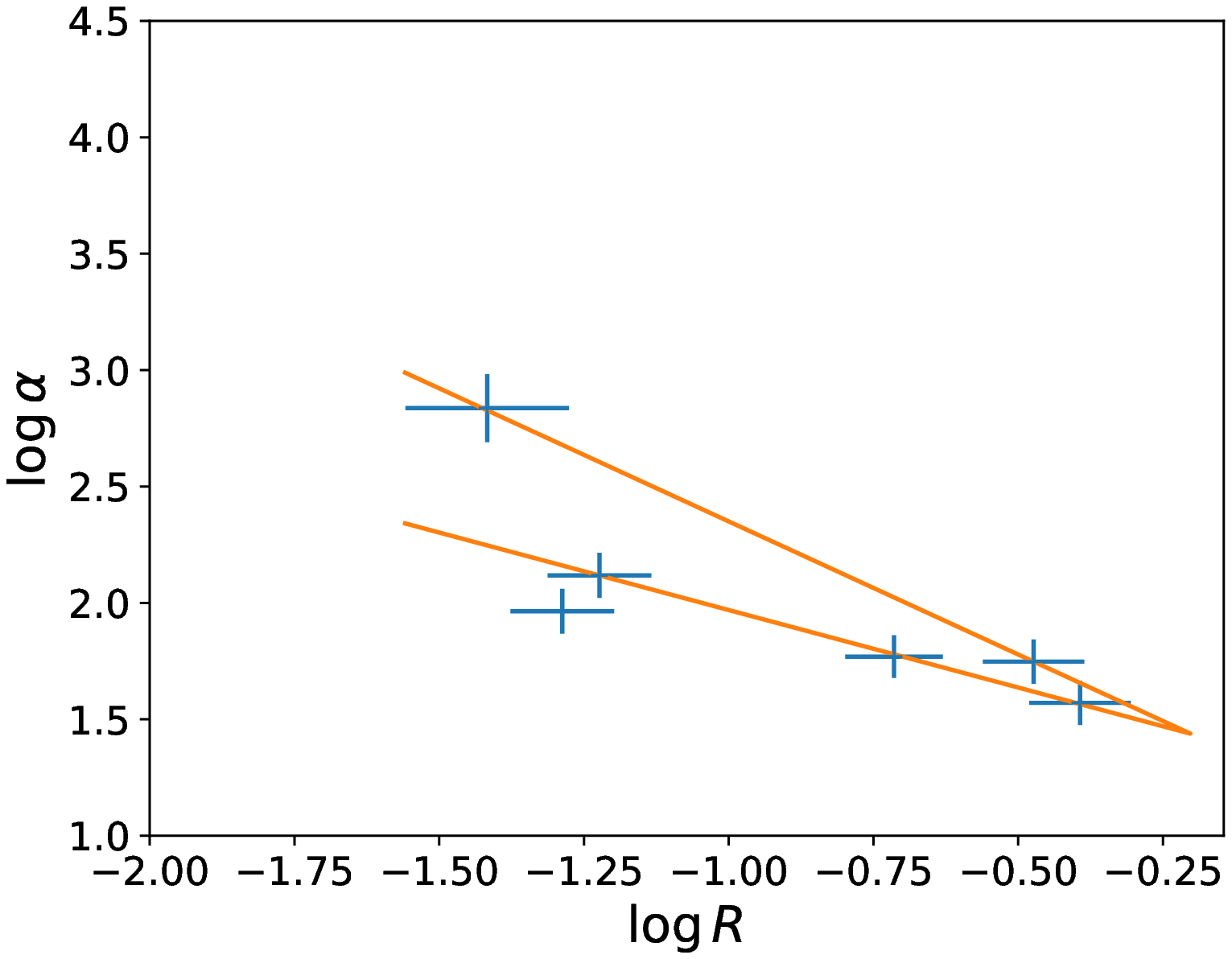}
    \end{center}
    \caption{The burst rate as a function of accretion rate ({\it left column}) and $\alpha$ as a function of burst rate ({\it right column}) for each source analysed in this paper. The measured values from MINBAR are plotted with Poisson errors assumed ({\it blue symbols}). The model fits described in \secref{sec:fits} are overplotted ({\it orange lines}). The grey areas in the first column of the plot for 4U~1636--536 indicate the accretion rates for which mHz QPOs have been detected, as reported by  \citep{art-2015-lyu-etal}.} 
    \label{fig:ras}
  \end{figure*}
}

\newcommand{\figmc}{%
  \begin{figure}
    \begin{center}
      \includegraphics[width=\columnwidth]{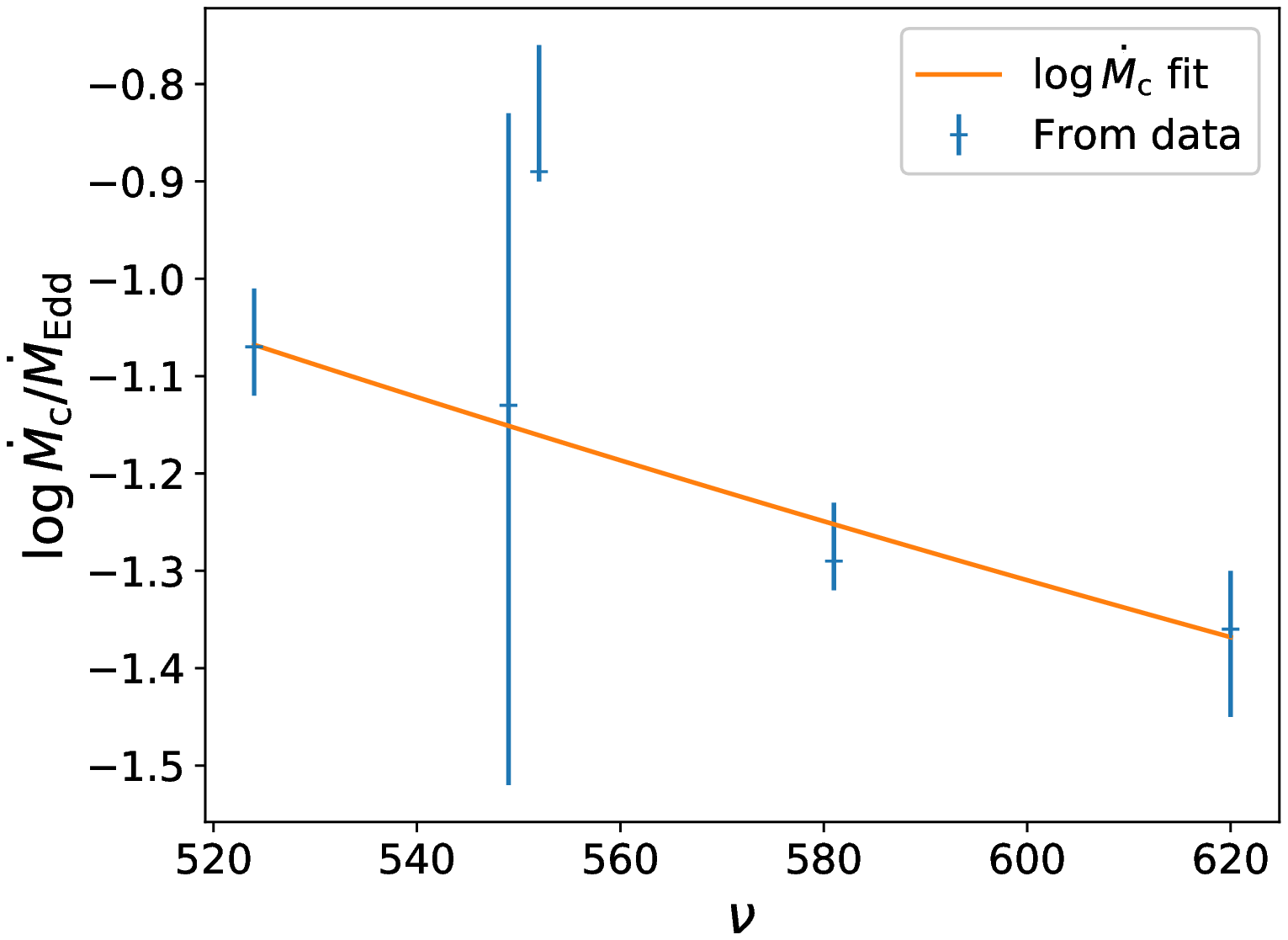}
    \end{center}
    \caption{The $\mc$ from the fit to the data in \figref{fig:ras}.  The line is a fit to these critical values.}
    \label{fig:mdot}
  \end{figure}
}

\newcommand{\figlat}{%
  \begin{figure}
    \begin{center}
      \includegraphics[width=0.9\columnwidth]{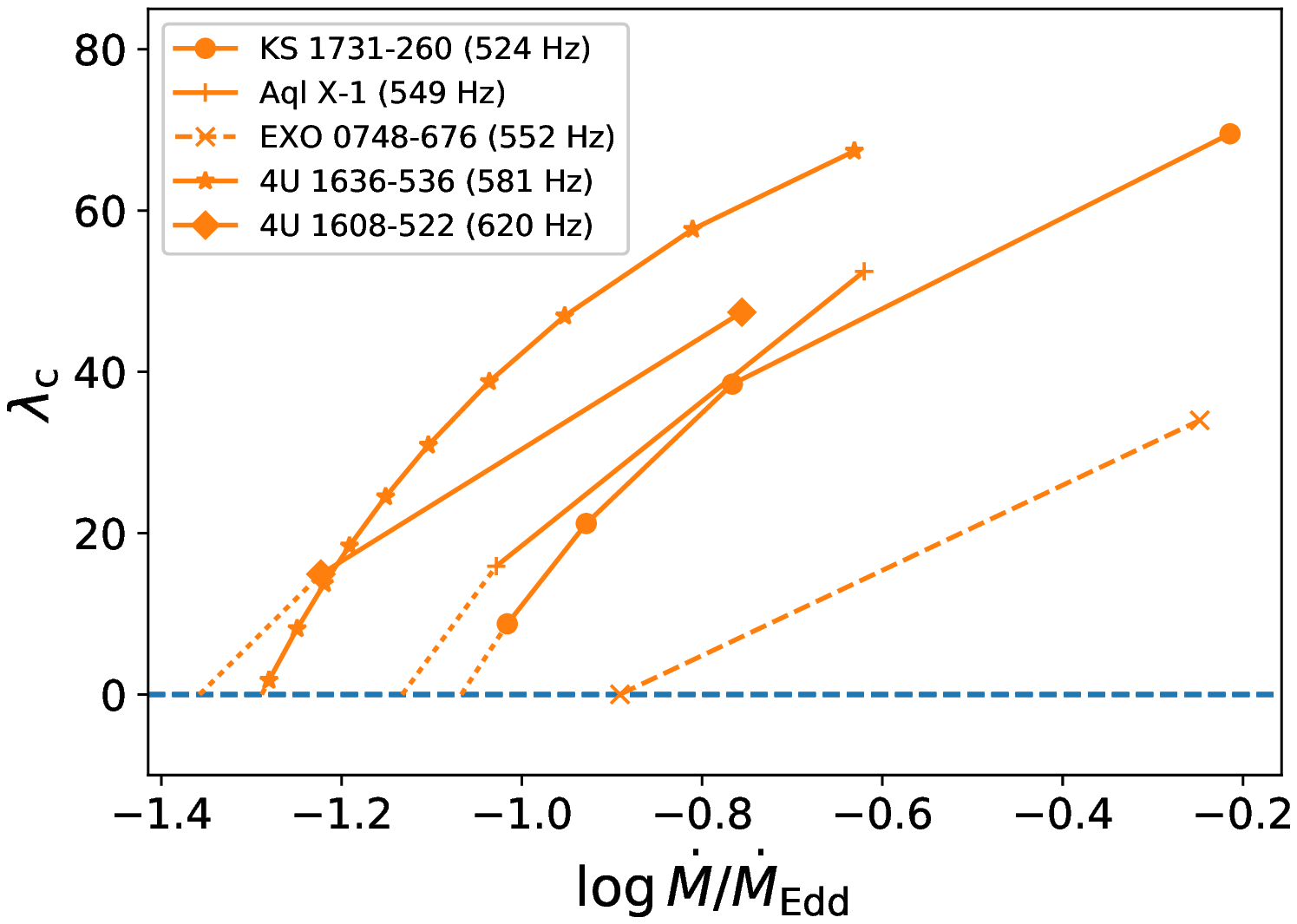}
    \end{center}
    \caption{The estimated ignition latitude based on \eqr{equ:area}.  The \textsl{blue dashed line} indicates the equator, where ignition is expected to take place on the first branch.  The \textsl{solid orange lines} indicate the ignition latitude on the second branch.  The \textsl{dotted segments} show the accretion rate at which the second branch starts.  The fact that higher spin correlates with an early wake of stabilisation can be seen.  The line for EXO~0748--676 is \textsl{dashed} to indicate the uncertainty on the accretion rates for this source.}
    \label{fig:lat}
  \end{figure}
}

\newcommand{\tabfit}{
  \begin{table*}
    \begin{center}
      \begin{tabular}{|c|c|c|c|c|c|c|c|c|c|}
        \etab{source}{$\nu$ [Hz]} & $\rmo$ & $\rmt$ & $\amo$ & $\amt$ & $\mc$ & $\rc$ & $\ac$ & Res$_{\lr}$ & Res$_{\lal}$\\
        \hline
        \etab{\ks{}}{524}  & \fit{0.73}{0.27}{0.27} & \fit{-1.02}{0.21}{0.24} & \fit{-0.61}{0.34}{0.39} & \fit{-1.99}{0.20}{0.25} & \fit{-1.07}{0.06}{0.05} & \fit{-0.34}{0.05}{0.05} & \fit{1.59}{0.08}{0.10} & 0.11 & 0.04 \nlbreak
        \etab{\aqu{}}{549} & \fit{1.11}{2.94}{0.60} & \fit{-2.17}{1.29}{2.98} & \fit{-0.61}{0.41}{0.37} & \fit{-1.23}{0.34}{0.40} & \fit{-1.13}{0.30}{0.39} & \fit{-0.41}{0.20}{0.17} & \fit{2.39}{0.21}{0.33} & 0.01 & 0.04 \nlbreak
        \etab{\exo{}}{552} & \fit{2.50}{0.43}{0.55} & \fit{-0.68}{0.15}{0.67} & \fit{-0.36}{0.24}{0.36} & \fit{-1.17}{0.51}{0.45} & \fit{-0.89}{0.13}{0.01} & \fit{-0.26}{0.25}{0.04} & \fit{2.15}{0.08}{0.25} & 0.28 & 0.03 \nlbreak
        \etab{\us{}}{581}  & \fit{1.57}{1.02}{1.02} & \fit{-1.62}{0.22}{0.27} & \fit{-0.46}{0.37}{0.52} & \fit{-1.50}{0.13}{0.15} & \fit{-1.29}{0.06}{0.03} & \fit{-0.25}{0.05}{0.06} & \fit{1.58}{0.08}{0.08} & 0.10 & 0.11 \nlbreak
        \etab{\ue{}}{620}  & \fit{2.85}{0.87}{0.82} & \fit{-2.01}{0.72}{0.54} & \fit{-0.67}{0.23}{0.24} & \fit{-1.14}{0.24}{0.28} & \fit{-1.36}{0.06}{0.09} & \fit{-0.20}{0.10}{0.17} & \fit{1.44}{0.20}{0.16} & 0.48 & 0.08 \nlbreak
      \end{tabular}
    \end{center}
    \caption{Source name, spin and results of the fit of \eqrs{equ:raone} - \eqref{equ:ratwo} following the procedure described in \secref{sec:fits}. The last two columns are the root mean square of the fit residuals of $\lr$ vs $\lm$ and $\lal$ vs $\lr$. The values for the second branch of EXO 0748--676 should be treated with care because we only have one data point on this branch.}
    \label{tab:data}
  \end{table*}
}

\newcommand{\tabsour}{
  \begin{table*}
    \begin{center}
      \begin{threeparttable}
        \begin{tabular}{|c|c|c|c|c|c|c|c|c|c|c|c|}
          Source & $\nu$ [Hz] & $P_{\rm orb} [hr] $ & $n_b$ & Exposure (Ms) & $c_{\rm bol}$ & $\xi_p/\xi_b$ \\
          \hline
          \ks{}~$^{a}$  & 524~$^{f}$ & ?~$^{k}$     & 366 & 18.8 & $1.62\pm0.13$ & 0.809/0.898\\
          \aqu{}~$^{b}$ & 549~$^{g}$ & 18.9~$^{l}$  &  96 & 7.49 & $1.65\pm0.10$ & 0.809/0.898 \\
          \exo{}~$^{c}$ & 552~$^{h}$ & 3.82~$^{m}$  & 357 & 22.9 & $1.6\pm0.3$   & 7.27/1.639\\
          \us{}~$^{d}$  & 581~$^{i}$ & 3.8~$^{n}$   & 664 & 11.4 & $1.51\pm0.12$ & 0.809/0.898\\
          \ue{}~$^{e}$  & 620~$^{j}$ & 12.9?~$^{o}$ & 145 & 11.7 & $1.59\pm0.13$ & 0.809/0.898\\
          \hline
        \end{tabular}
        \begin{tablenotes}
        \item[$^{a}$] \citet{atel-1989-suny-kwant}. $^{b}$ \citet{art-1976-lew-etal}.
          $^{c}$ \citet{atel-1985-par-etal}.
        \item[$^{d}$] \citet{atel-1976-swank-etal}. $^{e}$ \citet{art-1976-beli-con-eva}.
        \item[$^{f}$] \citet{art-1997-smi-etal}. $^{g}$ \citet{art-1998-zhang-all}.
          $^{h}$ \citet{atel-2009-gal-etal}.
        \item[$^{i}$] \citet{art-1998-stroh-etal}.
          $^{j}$ \citet{art-2002-muno-etal}.
        \item[$^{k}$] Unknown, but \citet{art-2000-muno-etal} suggest that it could be $\sim$ 1 hr.
        \item[$^{l}$] \citet{art-1991-cheva-ilo}.
          $^{m}$ \citet{art-1986-cramp-etal}.
          $^{n}$ \citet{art-1990-par-etal}.
          $^{o}$ \citet{art-2002-wach-etal}.
        \end{tablenotes}
      \end{threeparttable}
    \end{center}
    \begin{center}
      \caption{Sources adopted for analysis in this paper, along with the relevant analysis parameters: spin frequency, $\nu$; orbital period, $P_{\rm orb}$; number of bursts, $n_b$; total exposure; bolometric correction, $c_{\rm bol}$, and anisotropy correction, $\xi_p/\xi_b$ \citep[see text for details and][]{rev-arx-2020-gall-etal}.}
      \label{tab:sources}
    \end{center}
  \end{table*}
}

\title[Burst nuclear efficiency vs accretion rate]{The efficiency
  of nuclear burning during thermonuclear (Type I) bursts as a function of accretion rate}

\author[Y.~Cavecchi et al.]{%
Y. Cavecchi$^{1-3}$\thanks{E-mail:y.cavecchi@soton.ac.uk}, D. K. Galloway$^{4,3,5}$, A. J. Goodwin$^{4,3}$, Z. Johnston$^{6,3}$,   A. Heger$^{4,3,5,7}$\\%
$^1$Mathematical Sciences and STAG Research Centre,  University of Southampton, SO17 1BJ, UK\\%
$^2$INAF, Istituto di Astrofisica e Planetologie Spaziali, Roma 00133, Italy\\%
$^3$Joint Institute for Nuclear Astrophysics --- Center for the Evolution of the Elements (JINA-CEE)\\%
$^4$School of Physics and Astronomy, Monash University, Clayton, VIC 3800, Australia\\%
$^5$OzGrav-Monash --- School of Physics and Astronomy, Monash University, VIC 3800, Australia\\%
$^6$Department of Physics and Astronomy, Michigan State University, East Lansing, MI 48824, USA\\%
$^7$Center of Excellence for Astrophysics in Three Dimensions (ASTRO-3D), Australia
}

\begin{document}

\pagerange{\pageref{firstpage}--\pageref{lastpage}} \pubyear{}
\maketitle{}
\label{firstpage}

\begin{abstract}
  We measured the thermonuclear burning efficiency as a function of accretion rate for the Type I X-ray bursts of five low-mass X-ray binary systems. We chose sources with measured neutron star spins and a substantial population of bursts from a large observational sample. The general trend for the burst rate is qualitatively the same for all sources; the burst rate first increases with the accretion rate up to a maximum, above which the burst rate declines, despite the increasing accretion rate.  At higher accretion rates, when the burst rate decreases, the $\alpha$-value (the ratio of accretion energy and burst energy) increases by up to a factor of $10$ above that in the rising burst rate regime.  These observations are contrary to the predictions of 1D numerical models, but can be explained as the consequence of a zone of stable burning on the neutron star surface, which expands with increasing accretion rate.  The stable burning also ``pollutes'' the unstable burning layer with ashes, contributing to the change in burst properties measured in the falling burst rate regime. We find that the mass accretion rate at which the burst rate begins to decrease is anti-correlated with the spin of the neutron star. We conclude that the neutron star spin is a key factor,  moderating the nuclear burning stability, via the local accretion rate and fuel composition over the star.
\end{abstract}


\section{Introduction}

Type I X-ray bursts are nuclear flashes that take place on the surface of neutron stars in low mass X-ray binary systems.  These systems harbour a companion with a mass smaller than, or comparable to, that of the neutron star.  The companion fills its Roche Lobe and thus loses matter to the neutron star, feeding an accretion disc.  Matter from the disc subsequently falls onto the neutron star and spreads across its surface.  This fresh material, coming from the outer layers of the companion, can contain hydrogen, helium and metals that all can burn through nuclear reactions \citep{rev-1993-lew-par-taa}.

When the right conditions of pressure and temperature are met, the cooling processes cannot compensate for the heat released by the reactions and the burning becomes unstable, leading to a thermonuclear (Type I) X-ray burst.  Several authors have estimated the conditions for instability for different fuel compositions and accretion rates based on theoretical considerations \citep[see, e.g.,][]{rev-1998-bild, art-2003-nara-heyl, art-2003-cum, art-2004-woos-etal, art-2007-coop-nara-a, art-2007-heg-cumm-gal-woos, art-2011-keek-heger, art-2016-lampe-heg-gal, art-2016-kee-heg}.  Some of the important predicted observables are the burst rate and the accretion at which the bursts should stop (the burning becoming stable). In particular, theory consistently predicts that the burst rate should increase with increasing \emph{local} accretion rate, since the conditions for instability should be met earlier.  Above some critical accretion rate, the bursts should cease, because the burning becomes stable and can proceed sufficiently fast to consume all the freshly accreted fuel.

Observations, however, show that this is not the case. Instead, many sources show a critical accretion rate above which the burst rate begins to decrease \citep{art-1988-par-pen-lew, art-2003-corne-etal, rev-2008-gal-mun-hart-psal-chak, art-arx-2018-greb-chelo}. The rate may continue to decrease over a substantial range of accretion rates before the bursts disappear completely. For example, see \figref{fig:ras}.  There are indications that this critical accretion rate becomes lower as the spin of the neutron star increases \citep{art-2018-gal-etal}.  Another marked discrepancy with theory is the fact that the disappearance of the bursts happens at $\sim 0.1$ times the Eddington accretion rate and not at the Eddington accretion rate, as predicted \citep[e.g.][]{rev-1998-bild}.  Part of these discrepancies may come from the uncertainties in the nuclear reaction rates, for example of the key $3\alpha$ rate, but even within the allowed experimental uncertainties theory cannot meet the observations (see for example \citealt{art-2014-keek-cyb-heger, art-2016-cyb-etal}; see \citealt{rev-2018-meis-etal} for a review).

Proposed explanations for the decreasing burst rate include geometrical effects that would change the effective {\it local} accretion rate, and thus the burst rate, whereas the {\it global} accretion rate continues to increase \citep{rev-1998-bild}. Such an explanation, however, would lead to an even higher discrepancy with the predicted accretion rate of final stabilisation. Another possible solution comes from the effects of extra heat sources near the burning layer, which may both stabilise or accelerate the occurrence of bursts (Type I and longer ones).  This direction is very promising, since a similar heat source deeper in the crust seems to be needed also from observations of the cooling curves of neutron stars \citep{art-2009-brow-cum, rev-2017-wij-dege-page}, but it introduces the problem of the origin of this heat: the so called \emph{shallow heating problem}.  A possible origin could be kinetic energy dissipated at the spreading layer where disc matter touches the neutron star \citep{art-1999-ino-suny}.  Detailed simulations are needed, however, to quantify this effect.  Another possible source is further nuclear reactions or kinetic energy dissipative effects near the burning layer \citep[e.g.][]{art-2004-brow, art-2005-coop-nara, art-2006-cum-mac-tzan-page, art-2007-piro-bild, art-2009-kee-lang-zand, art-2011-keek-heger}. Furthermore, even the heat from previous bursts as well as "chemical inertia" \citep{art-2004-woos-etal} is also very important, as demonstrated by the train of simulations by \citet{art-2018-johnst-etal} modelling the bursts from the accretion-powered millisecond pulsar SAX~J1808.4--3658.

\tabsour{}

Some possible solutions have also been studied using one-zone or one-dimensional (1D, radial) models.  \citet{art-2007-piro-bild} and \citet{art-2009-kee-lang-zand} explored the effects of composition and mixing in 1D analytical and numerical calculations, respectively, finding encouraging results. Another explanation put forward for the decreasing burst rate is the delayed detonation regime of \citet{art-2003-nara-heyl}.  This regime, however, has not been reproduced by other softwares, notably \kep{} \citep[see also the notes in \citealt{art-2007-coop-nara-b}]{art-2014-keek-cyb-heger} nor confirmed by experimental observations\footnote{While it is true that \citet{art-2016-lampe-heg-gal} saw a qualitatively similar behaviour in their \kep{} simulations, this was within a very narrow range of mass accretion rate.}. An additional complication is that differences in local effective gravity with latitude, due to the rotation of the star, may change the stabilisation conditions and lead to the observed behaviour, as suggested by \citet{art-2007-coop-nara-a}.  The dependence on gravity, however, is weak and its change with latitude is at most a few percent, implying that this explanation might only work for fast rotators with $\nu>\sim600\,\mathrm{Hz}$.

A further piece of evidence supporting the role of stable burning in explaining the decreasing burst rate, is the presence of milli-Hertz quasi-periodic oscillations (mHz QPOs) in some burst observations.  These QPOs were first detected in 4U~1636--536 by \citet{art-2001-rev-etal}, who associated them with some kind of nuclear burning. \citet{art-2007-heg-cum-woos} proposed that they arise from ``marginally stable'' burning, which is present at accretion rates close to the boundary for stability, as first suggested by \citet{Pac83}.  The plausibility of this interpretation was further reinforced by \citet{art-2008-alta-all-a}, who found that in \us{} the QPOs disappeared before the onset of Type I bursts and by \citet{art-2012-lin-etal}, who observed the transition of regular bursts into mHz QPOs for the source IGR~J17480--2446 (in the globular cluster Terzan 5).  In \us{}, however, this transition does not seem to take place at the theoretically expected mass accretion rate. \citet{art-2009-kee-lang-zand} suggested that mixing of ashes with fresh accreted fuel could explain the presence of mHz QPOs at different accretion rates, and their semi-analytical models agree relatively well with the observations of \citet{art-2008-alta-all-a}.  However, \citet{art-2014-keek-cyb-heger}, who ran various simulation changing composition and the reaction rates within experimental errors, were not able to obtain mHz QPOs at the observed accretion rates for any ``reasonable'' fuel composition or reaction rates.

In \citet{art-2017-cavecchi-etal} we used simplified analytical relations to move beyond 1D considerations.  We showed how different conditions on the surface of the neutron star, in particular a variation of composition along latitude, due perhaps to rotationally-induced mixing, could lead to early stabilisation of some parts of the star, while other parts still contribute to the bursting behaviour.  Using our parametrisation we showed which specific configurations could resolve the apparent dichotomy between one dimensional theoretical predictions and the observations, and how to connect observed parameters with theoretical ones.  In a parallel study, we showed that the accretion rate at which the burst rate reached a maximum depends on the spin rate of the bursting neutron star \cite[]{art-2018-gal-etal}.  Specifically, the accretion rate at burst-rate peak (where it can be measured) {\it decreases} with increasing spin rate; this effect is the opposite of what would be expected if rotational shear was contributing additional heat resulting in stabilisation of the thermonuclear burning \citep{art-1999-ino-suny}.

In this paper, we present our study of the ample sample of available burst data for indications in support of our picture: in \secref{sec:obs} we describe the sources and the observations we used, in \secref{sec:results} we present our results and in \secref{sec:disc} we discuss our conclusions.

\section{Observations and analysis}
\label{sec:obs}

\subsection{Source data}

Our analysis is based on the public data available in the Multi-INstrument Burst ARchive (\minb{}; \citealt{rev-arx-2020-gall-etal}). \minb{} includes more than 7000 bursts detected by past and present X-ray instruments onboard \sat{BeppoSax}, \sat{RXTE} and \sat{INTEGRAL}.  The available data provide a homogeneous analysis of both the bursts and the host observations, with coverage that extends over many years and, in some cases, over a wide range of accretion rates.  The specific parameters we used for the bursts were the estimated accretion rate and bolometric fluence ({\tt gamma}, {\tt bfluen}) and, for the observations, the estimated accretion rate and duration ({\tt gamma}, {\tt exp}).  See the \minb{} paper \citep{rev-arx-2020-gall-etal} for a complete description of how these parameters are derived.

We report results for the sources listed in \tabref{tab:sources}.  We have chosen these sources for two reasons: first, these sources show a decrease in burst rate while the mass accretion rate still increases and, second, their spin frequency has been measured from the detection of burst oscillations (see references in \tabref{tab:sources}).  The sources are: \ks{} with $\nu=524\,\mathrm{Hz}$; \aqu{} with $\nu=549\,\mathrm{Hz}$; \exo{} with $\nu=552\,\mathrm{Hz}$; \us{} with $\nu=581\,\mathrm{Hz}$ and \ue{} with  $\nu=620\,\mathrm{Hz}$.  These are transient sources with a range of orbital periods: possibly $\sim1\,\mathrm{hr}$, $19\,\mathrm{hr}$, $3.82\,\mathrm{hr}$, $3.8\,\mathrm{hr}$, and $12.89\,\mathrm{hr}$ (see references in \tabref{tab:sources}).

For each source, we calculated the burst rate as a function of the inferred accretion rate, following \cite{art-2018-gal-etal}.  The estimated accretion rate for each observation in \minb{} is given by $\gamma$ ({\tt gamma}), the estimated bolometric persistent flux, rescaled by the measured Eddington flux (using photospheric radius-expansion bursts).  The $\gamma$-factors also take into account the expected effects of anisotropy of the burst and of the persistent emission.
We grouped the bursts into bins of $25$ or $50$, choosing the larger value for sources with more bursts over all, and estimated the rate within each bin as 
\begin{equation}
    R_i = \frac{n_i}{\sum_j \Delta t_{j,i}}
\end{equation}
where $n_i$ is the number of bursts with $\gamma$ values falling in bin $i$, and $\sum_j\Delta t_{j,i}$ is the total exposure for all the observations with $\gamma$ values falling in bin $i$.  We excluded so-called ``short recurrence-time'' bursts by excluding those events occurring $<30$~min before the previous one. These events are thought to arise from ignition of residual fuel from the previous event and are not representative of the conventional ignition conditions we study here \citep{art-2010-keek-etal}.  We estimated the uncertainties in each bin based on the expected Poisson variance of the number of bursts falling in each bin.

\figres{}
\tabfit{}

The efficiency of burning is usually estimated via the parameter $\alpha$, that is defined as the ratio between the average energy released by the persistent emission between the bursts and the energy released during the bursts:

\begin{equation}
  \alpha = \frac{E_{\rm pers}}{E_{\rm bur}}= \frac{Q_{\rm grav}}{Q_{\rm nuc,burst}}
  \label{equ:alpha}
\end{equation}
where $Q_{\rm grav} = c^2z/(1+z) \approx GM_{\rm NS}/R_{\rm NS}$ is the energy released through accretion and $Q_{\rm nuc,burst}$ is the nuclear burning energy released during the burst.  $\alpha$ is higher when the energy released during the bursts becomes smaller.  For example, the efficiency can be reduced when the nuclear reactions produce less stable nuclei, or when some fraction of the accreted matter burns stably between bursts and hence less remains available during the unstable flashes.

The $\alpha$-value is usually estimated by integrating the flux between the bursts and then dividing it by the integrated flux (the fluence) of the bursts.  However, since both the persistent and burst emission are affected by the geometry of the source \cite[e.g.][]{he16}, the measured value will be biased compared to the theoretical value $\alpha$ defined above.
For the purposes of this study, we  estimated the average ratio of persistent to burst luminosity, $\alpha_i$, within each bin, as follows.  We extracted from \minb{} the bolometric fluence $E_b$ for each burst in the sample, where available.  Fluence measurements are not available for some bursts, including those with incomplete data, or those detected with {\it INTEGRAL}/JEM-X.  For those missing fluences we assumed values equal to the average of those in the same bin with measured values.  We then estimated the alpha value as
\begin{equation}
    \alpha_i = \frac{c_{\rm bol}\sum_j F_{p,j,i}\Delta t_{j,i}}{\sum_kE_{b,k,i}} \frac{\xi_p}{\xi_b}
\end{equation}
where $F_{p,j,i}$ and $\Delta t_{j,i}$ are the $3$--$25\,\mathrm{ke\!V}$ persistent X-ray flux and duration of each observation falling into bin $i$, and $E_{b,k,i}$ is the estimated bolometric fluence of each burst falling in bin $i$.  As the persistent flux measurements are made within a restricted energy range of $3$--$25\,\mathrm{ke\!V}$, we multiply by a bolometric correction factor $c_{\rm bol}$, which is adopted for each source from \minb{}.  We also correct for the expected anisotropy of the burst and persistent emission, by multiplying by the ratio $\xi_p/\xi_b$, as adopted for the specific source type \citep[dipper/non-dipper, see][based on \citealt{he16}]{rev-arx-2020-gall-etal}.

\subsection{Burst rate fitting}
\label{sec:fits}

We call first and second \emph{branch} the regimes when the burst rate first increases and then decreases.  Following \cite{art-2017-cavecchi-etal}, we seek to identify the conditions for each system when the second branch begins. We refer to these inflection points as the critical mass accretion rate, burst rate and $\alpha$: $\dot{M}_{\rm{c}}$, $R_{\rm{c}}$ and $\alpha_{\rm{c}}$, respectively. We fit the log base ten of the burst rate versus the log base ten of the accretion rate.

We also fit the log base ten of $\alpha$ versus the log burst rate. The fits are linear regressions in log space, one for each branch. Motivated by the idea that different local composition should affect the burning and thus both burst rate and $\alpha$ \citep{art-2017-cavecchi-etal} and noticing similar inflection points in the observations for the measured $\alpha$ values, the fits are performed simultaneously, enforcing that the critical burst rate $R_{\rm{c}}$ corresponding to the turning point of the two lines on the $\dot{M}-R$ plane is the same on the $R-\alpha$ plane.  Effectively, we assume that
\begin{align}
  \label{equ:raone}
  \lr &= \rmo \lm + \rno\\
  \lr &= \rmt \lm + \rnt\\
  \intertext{and}
  \lal &= \amo \lr + \ano\\
  \lal &= \amt \lr + \ant
  \label{equ:ratwo}
\end{align}
with $\rno = \rc - \rmo \mc$, $\rnt = \rc - \rmt \mc$, $\ano = \ac - \amo \rc$, $\ant = \ac - \amt \rc$. We fit for $\rmo$, $\rmt$, $\mc$, $\rc$, $\ac$, $\amo$ and $\amt$ for each source.

\section{Results}
\label{sec:results}

In \figref{fig:ras} we show the variation in burst rate and efficiency.  Each row corresponds to one of our sources, in order of increasing spin from top to bottom.  In the first column we show the burst rate as a function of the accretion rate.  The behaviour observed by \citet{art-2003-corne-etal} is apparent: the burst rate divides into two \emph{branches}, one is increasing and the other is decreasing.
We also show the best-fit curves from the model described in \secref{sec:fits}, using the fit parameters listed in \tabref{tab:data}.
The data is of sufficient quality for each source to constrain the accretion rate at which the burst rate peaks.  The only potential exception is EXO~0748--676, for which the second branch is insufficiently resolved to draw firm conclusions, with only one burst rate bin for accretion rates above the burst rate peak.

The second column of \figref{fig:ras} shows the variation of the measured $\alpha$-value as a function of burst rate and the corresponding model fits.  The increase of $\alpha$ with increasing accretion rate was noted before \citep[see e.g.][]{art-1988-par-pen-lew,art-2003-corne-etal}, but we have found that \emph{also $\alpha$ displays two branches}.  Where the burst rate is increasing, i.e. on the first branch, $\alpha$ is constant or slowly decreasing, meaning a higher fraction of the nuclear energy of the accreted material is released in the bursts.  On the second branch, where the burst rate decreases, all sources consistently display an increase in $\alpha$, implying less nuclear energy is released in the bursts.  Most importantly, the $\alpha$ from this second branch does not follow the $\alpha$ from the first branch, but it has a significantly steeper slope, indicating substantially smaller fraction of the nuclear energy being released by the bursts on this branch. 

We propose that this decrease of the relative ``bursting efficiency'' is mainly the result of two factors, which both arise from {\it stabilisation of the burning on a growing fraction of the star}, most probably around the equator.  First, the stable burning at the equator produces more ashes which may pollute the rest of the star and, second, because the equator contributes to the persistent emission, a smaller surface area of the star contributes to the burst energy, thus increasing $\alpha$ as per \eqr{equ:alpha}.

We also attempted to corroborate our explanation based on stable burning with the presence of mHz QPOs.  In their study of the mHz oscillations detected in \us{}, \citet{art-2015-lyu-etal} tabulated each {\it RXTE}/PCA observation in which mHz oscillations are present, observations which are also part of the \minb{} sample. Thus, in the first column of \figref{fig:ras} for \us{} we plotted grey areas at the accretion rates where the mHz QPOs have been detected. Unfortunately, no such information is available for the other sources in our study.

\section{Discussion and conclusions}
\label{sec:disc}

\subsection{A wake of stabilisation}

We measured the burst rate and $\alpha$ as a function of accretion rate for the five sources \ks{,} \aqu{,} \exo{,} \us{} and \ue{.}  These are shown in \figref{fig:ras}, where the sources are arranged by increasing spin from top to bottom.

The most striking feature is the division into two regimes, both for burst rate and $\alpha$, which we termed first and second branch.  The first branch approximately follows the predictions of 1D models, while the second does not.  It is remarkable that while on the first branch the burst $\alpha$ decreases slightly with accretion rate and burst rate, on the second branch $\alpha$ increases markedly with accretion rate, contrary to 1D predictions.

In \citet{art-2017-cavecchi-etal} we started from the premise that one could consider the surface of the neutron star as a series of 1D (radial) models located at different latitudes, each characterised by a different composition and thus different intrinsic burst rate and stabilisation critical local accretion rate.  As a corollary, a different local $\alpha$ is to be expected as well.  We then explained the decreasing burst rate on the second branch as the stabilisation of the highest-rate latitude and the progressive switching to the burst rate of the other latitudes.  Most likely, based on observed parameters and accretion theory, the equator has initially the fastest burst rate, but it is also the latitude that stabilises first \citep[see cases a or c of Figures 4 and 5 in][ and references therein]{art-2017-cavecchi-etal}.

To clarify terminology in the ensuing discussion, we distinguish between ``steady'' and ``stable'' burning.  We refer to \emph{steady burning} as the temperature- and density-insensitive burning of hydrogen by the $\beta$-limited CNO cycle; and to \emph{stable burning} for usual temperature- and density-dependent burning processes, such as the $3\alpha$-, $\alpha$p- and rp- processes, when encountered in a stable fashion as in 1D models at about local Eddington accretion rate \citep{art-2007-heg-cum-woos, Zam14}.  The stable helium burning of \citet{art-2016-kee-heg} would fall in the second regime, but is somewhat different and not relevant here.

On the first branch, the trend of increasing burst rate with increasing accretion rate is consistent with the predictions from the 1D models.  On the second branch, the growing role of steady burning as the accretion rate increases explains both why the burst rate does not \emph{apparently} follow 1D predictions, and why the measured $\alpha$-value increases so dramatically.  We propose the higher $\alpha$-values in the second branch can be understood as the combination of several effects.  First, $\alpha$ is indeed higher for lower burst rate (1D models), because the fuel composition at ignition is more H-poor due to the effects of steady burning over longer inter-burst intervals.  Second, the persistent emission increases due to the contribution from stable burning over the part of the surface that passed the stabilisation critical local accretion rate (2D effect).  Third, the bursts release less energy because a greater fraction of the accreted material has been burned stably (2D effect).  Fourth, the composition of the latitudes that drive the bursting on the second branch has an intrinsic higher $\alpha$ due to the composition at those latitudes (2D effect plus different mixing effects at different latitudes).

In support of our explanation, we note that for \us{,} mHz QPOs appear precisely where we expect stabilisation to be about to begin: on the first branch, right before the peak of the burst rate.  The mHz QPOs also remain present for almost all the second branch, where we expect the wake of stabilisation to be moving towards the poles.  In our interpretation, the increase of burst $\alpha$ with accretion rate on the second branch is the result of ignition of more polluted fuel towards the poles, while a growing region around the equator is transitioning to stable burning.  The mHz QPOs could be ignitions that take place at the latitude of the boundary between the stable and the unstable burning regions.  As long as bursts are observed, there should be such a boundary, which would explain the presence of mHz QPOs all the way down the second branch.  The different local conditions of mass accretion rate and composition at the different latitudes could then be the missing factor to link 1D simulations of mHz QPOs to observations.

\subsection{The latitude of stabilisation}

\figlat{}

One could make a simplifying assumption to measure the effects of stabilisation and locate the stabilisation latitude.  Let's assume that for a given accretion composition the local intrinsic $\alpha$ is just a function of burst rate, basically equivalent to saying that there is the same composition across the whole surface all the time.  If we say that the total burst energy in \eqr{equ:alpha} is proportional to the area burning unstably, we could write $E_{\rm bur} = \epsilon_{\rm burst} 4\pi R_*^2 (1 - \sin \lambda_{\rm c})$.  Here $\epsilon_{\rm burst}$ is the energy released per unit surface and we assume symmetry between the northern and southern hemispheres. $\lambda_{\rm c}$ is the lowest latitude where burning is unstable.  When the entire surface is burning unstably (as we expect in the first branch), $\lambda_{\rm c}=0$, corresponding to the equator.  When the burning stabilises on the equator, $\lambda_{\rm c}>0$ and we expect to measure an $\alpha$-value on the second branch given by:
\begin{equation}
  \alpha_{2} = \frac{\alpha_{1}(R)}{1 - \sin \lambda_{\rm{c}}}
  \label{equ:area}
\end{equation}
where $\alpha_1$ is given by \eqr{equ:alpha} and it is to be assumed a function of burst rate only, so that the scaling is given by the difference in the areas involved in stable and unstable burning.  In order to apply this formula to our data, we make use of the fitted relations derived in \secref{sec:obs}. Through those relations we map $\lm$ to $\lr$ and this to $\lal$ on the two branches.  We then invert \eqr{equ:area} to obtain the latitude of the wake of stabilisation from the ratio of the $\alpha$s on the two branches corresponding to the same burst rate.  This is plotted in \figref{fig:lat}.  \aqu{}, \exo{} and \ue{} seem to reach up to $\lambda_{\rm{c}} \sim 40$--$50$~degrees, whereas the other sources seem to go up to $\lambda_{\rm{c}}\sim60$ degrees.  This dichotomy can be due to observational biases or to intrinsic differences in the accreted composition in the various systems (see the discussion in \secref{sec:diff}).

This measure is just a first approximation, because it neglects the possible role of varying local fuel composition, that are difficult to include without extensive modelling and further fitting, that is beyond the scope of this paper.  First of all, the simple formula of \eqr{equ:area} neglects the effects of a wider spreading layer due to higher accretion rate \citep{art-2014-kaja-etal}, which is difficult to quantify.  As for which profile of composition to expect at various latitudes, we pose the following speculation.  If new material lands only around the equator, the Coriolis force will confine it up to a latitude that depends on the spin (see Eq.~23 of \citealt{art-2015-cavecchi-etal}) and it would be difficult to bring fresh material northward and southward, at least considering only 2D (Cavecchi \& Cumming, in prep.).  The higher latitude composition would therefore be closer to nuclear ashes.  However, the baroclinic instability, which was found to be  relevant during the bursts \citep{art-2019-cavecchi-spit}, could break the confinement.  Even so, the faster the rotation, the harder it could be to reach higher latitudes (see discussion in \citealt{art-2019-cavecchi-spit} about the extent of the instability effects).  Detailed simulations should also quantify better how to connect the global accretion rate to the local effective one at any given latitude.

\begin{figure}
  \begin{center}
    \includegraphics[width=\columnwidth]{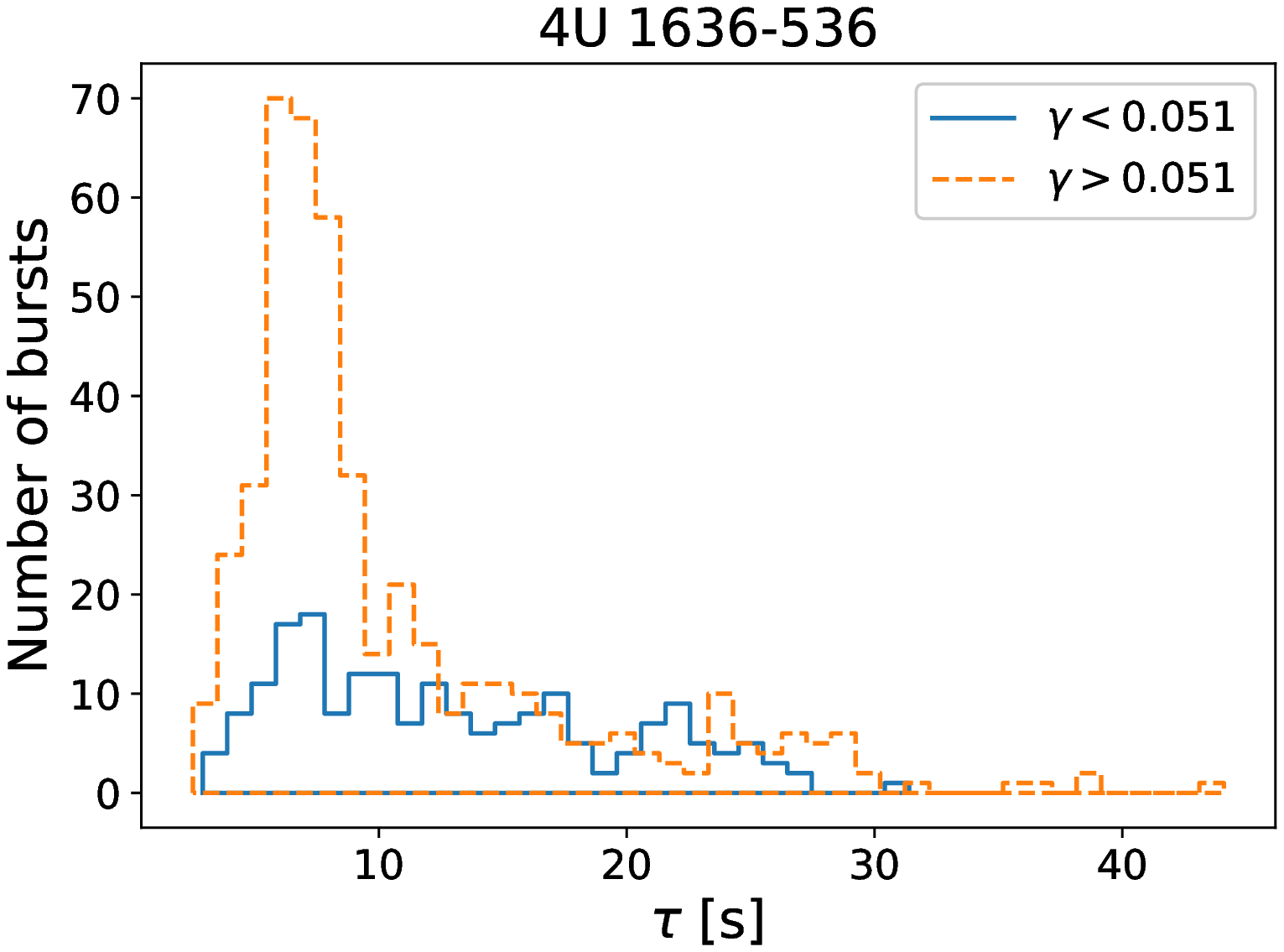}
    \caption{Distribution of burst timescale $\tau$ (the ratio of the burst fluence $E_b$ to the peak flux $F_{\rm pk}$) for bursts from 4U 1636--536 at low and high accretion rates. Note the strong preference for short duration bursts (similar to bursts from H-poor material) at high accretion rates, when long, frequent bursts would be expected.}
     \label{fig:tau}
  \end{center}
\end{figure}

Probably, the rise in latitude from \eqr{equ:area} is overestimated, if the local $\alpha$ at higher latitudes is \emph{intrinsically} higher than the equatorial one, as we suspect.  We can test for the effects of composition pollution by comparing the properties of the bursts in each accretion rate branch.  For example, in the case of \us{,} when the burst rate is rising we commonly see a wide range of burst timescales $\tau$ (the ratio of the burst fluence $E_b$ to the peak flux $F_{\rm pk}$), with the largest values indicative of mixed H/He burning consistent with the assumed accretion composition (\figref{fig:tau}).
At the highest accretion rates, numerical models would predict high burst rates providing less time for reducing the H-fraction below that of the accreted material.  Instead, we find increasingly {\it infrequent} bursts, with short timescales suggesting instead H-poor material.  If this material is due to pollution from lower latitudes, because of (partial) stable burning, it may actually be depleted in CNO material, similar to the effect observed in the short waiting time bursts of \citet{KH17}.  In this case, heating from nuclear burning could be reduced and burst rate decreased.  The effect of additional burning ashes that are neither H, He, nor CNO remains to be explored.

We performed a two-sided Kolmogorov-Smirnov test on the $\tau$ of the bursts of each of our sources, dividing them at the $\gamma$  of the break between the two branches ($\dot{M}_{\rm{c}}$ values in \tabref{tab:data}).  The corresponding p-values for the null hypothesis that the two samples of $\tau$ follow the same distribution are: \ks{:} $4.39\tent{-4}$, \aqu{:} $1.03\tent{-1}$, \exo{:} $4.77\tent{-1}$, \us{:} $1.32\tent{-9}$ and \ue{:} $1.37\tent{-4}$.  The values for \ks{,} \us{} and \ue{} indicate that there is little probability that the $\tau$ are from the same distribution, therefore supporting our picture.  The results for \aqu{} and \exo{} are less straightforward. In the case of \aqu{,} however, which can still be rejected at $\sim10\,\%$ confidence level, the bins on each branch are very close to the break, so that the burst may still have very similar composition.  In the case of \exo{} we only have one bin after the break because of very few bursts detected at the highest accretion rates, so that, again, a good fraction of the bursts in the second sample could be from when the composition is still not significantly different from the one setting the behaviour on the first branch.

We have compared the measured $\alpha$ to a library of models run in different conditions \citep{art-2018-johnst-etal, art-2019-goodw-heg-gal, art-2019-goodw-etal}.  We found that it is possible to obtain such high values by changing only the composition, if we combine low H fraction ($X\sim 0.2$) with also very low CNO metallicity ($Z_{\rm CNO}\sim 0.005$).  This would be consistent with mixing polewards (to the burst latitudes) material that underwent partial steady burning and CNO breakout close to the equator. These processes both increase He and destroy CNO, converting it to heavier elements that do no exhibit efficient cycles \citep{KH17}.  This may be combined with area `losses' due to stabilisation.  With more detailed data and models this approach could be applied more precisely and it would enable us to follow the evolution of the ignition on the surface at the various accretion rates.

\subsection{The role of different accreted composition and burning regimes}
\label{sec:diff}

Assuming that the conditions for stability are mainly dependent on the spin of the star, we have fitted a linear relation between the critical accretion rate $\mc$ and the star spin $\log \nu$ \citep[see \figref{fig:mdot} and ][]{art-2017-cavecchi-etal}.  The figure shows that the critical $\lm$s do indeed align very well.  The only outlier is \exo{.} 
\exo{} is a ``dipper'', which indicates a high system inclination, such that the outer edge of the accretion disk partially obscures the persistent (and burst) emission from the neutron star.  We attempted to take the high inclination into account by choosing specific anisotropy factors for this source, as noted in \tabref{tab:sources}.  The anisotropy factors shift the value of $\lm$ and therefore also of $\mc$: it seems plausible that these anisotropy factors could be exaggerated.  Nonetheless, the other data points do point to a common behaviour for the other sources.  Excluding \exo{}, the fit yielded the values: $\mc = -4.194 \log \nu + 10.339$. 

This inverse relation between $\mc$ and $\nu$ is what explains also why some sources display only the first or the second branch \citep{art-2003-corne-etal}. If they are rotating too slow, they may not experience sufficiently high accretion rate to reach their $\mc$ or, vice versa, if they spin too fast even a low accretion rate could be above their $\mc$ \citep{art-2017-cavecchi-etal}.

One caveat is that to be able to fit $\mc$ versus $\log \nu$ we are implicitly assuming that all these sources accrete the same composition.  Some of the discrepancies could be indeed related to (slight, but effective) differences in the composition of the companions, such as metallicity.  Comparing the plots in \figref{fig:ras} one may be tempted to qualitatively combine \aqu{} and \ue{} \citep[orbits of 19 hr and 12.89 hr,][]{art-1991-cheva-ilo, art-2002-wach-etal} in one category and \ks{,} \exo{} and \us{} \citep[orbits of $\sim$ 1 hr, 3.82 hr and 3.8 hr,][]{art-2000-muno-etal, art-1986-cramp-etal, art-1990-par-etal} in another group.  While the orbital periods may indicate a similarity in the binary evolution stage and therefore the composition of the companion, which may provide an initial confirmation of this grouping, our analysis is too limited at the moment to support this conclusion.

\figmc{}

The effect of the accreted composition was clear when we also analysed two additional sources: 4U 1702-429, $\nu=329\,\mathrm{Hz}$ \citep{art-1977-mars-li-rappa, art-1999-mark-stroh-etal}, and 4U 1728-34, $\nu=364\,\mathrm{Hz}$ \citep{art-1976-hoff-lew-etal, art-1996-stro-etal}.  These two sources are thought to be pure helium accretors in short orbit systems \citep{art-2007-zand-jon-markw, art-2010-misa-gall-coop, art-2010-gall-etal, art-2020-vinc-cavecchi-etal}, whereas the others are thought to accrete a composition which is richer in hydrogen. These two sources display a different behaviour from the others.  The burst rate increases constantly with accretion rate, while the $\alpha$ parameter remains around the same values, $\sim1.60$ - $2.3$, showing a slight increase with burst rate and, thus, accretion rate.  This difference is not in contrast to our interpretation, since the difference in accretion composition is expected to lead to different parameters for the value of the critical accretion rate of the turn over \citep[see][]{art-2017-cavecchi-etal}.

\subsection{Outlook}

A more comprehensive examination of the accretion rates at which mHz QPOs occur in the sources other than \us{} may help to add further evidence to support the link between stable burning and the second branch, where the burst rate falls and the $\alpha$-value rises. There is extensive archival data available that can address this question.  Furthermore, upcoming satellites like \textit{eXTP} \citep{art-2018-zhang-etal} or \textit{STROBE-X} \citep{art-arx-2019-ray-etal} have the instrumental capabilities to deliver exquisite data of Type I X-ray bursts and they will target accreting neutron stars which are also bursters for their core science cases.  We suggest that a long monitoring campaign, that would track the evolution of these and other sources with mass accretion rate, would be key to clarify the 2D nature of burning on accreting neutron stars, by providing more resolved data bins for better fits of the branching of burst rate and $\alpha$ value.  In the meantime, more accurate numerical modelling of the fluid distribution over the surface will provide a more precise model of the different ignition conditions on the star.  Gathering this information will make it possible to resolve transitions between different burning regimes and clarify whether the second branch evolution is due to a different burning regime which gives decreasing burst rate or if there is a wake of stabilisation and ignition in matter with different surface composition.

\section*{Data Availability Statement}
This research utilizes analysis results from the Multi-INstrument Burst ARchive (MINBAR) \citep[available \href{https://doi.org/10.26180/5e4a697d9b8b6}{here},][]{rev-arx-2020-gall-etal}.

\section*{Acknowledgements}
YC thanks the Monash Centre for Astrophysics (MoCA) for hospitality under the Distinguished Visitor program. This work was supported in part by the National Science Foundation under Grant No.\ PHY-1430152 (JINA Center for the Evolution of the Elements). MINBAR has benefited from support by the Australian Academy of Science's Scientific Visits to Europe program, and the Australian Research Council's Discovery Projects (project DP0880369) and Future Fellowship (project FT0991598) schemes. The MINBAR project has also received funding from the European Union's Horizon 2020 Programme under the AHEAD project (grant agreement no. 654215).  Parts of this research were conducted by the   Australian Research Council Centre of Excellence for Gravitational Wave Discovery (OzGrav), through project number CE170100004. AH has been supported in part by the Australian Research Council Centre of Excellence for All Sky Astrophysics in 3 Dimensions (ASTRO 3D), through project number CE170100013.

\bibliographystyle{mnras}
\bibliography{ms}

\begin{thebibliography}{}
\makeatletter
\relax
\def\mn@urlcharsother{\let\do\@makeother \do\$\do\&\do\#\do\^\do\_\do\%\do\~}
\def\mn@doi{\begingroup\mn@urlcharsother \@ifnextchar [ {\mn@doi@}
  {\mn@doi@[]}}
\def\mn@doi@[#1]#2{\def\@tempa{#1}\ifx\@tempa\@empty \href
  {http://dx.doi.org/#2} {doi:#2}\else \href {http://dx.doi.org/#2} {#1}\fi
  \endgroup}
\def\mn@eprint#1#2{\mn@eprint@#1:#2::\@nil}
\def\mn@eprint@arXiv#1{\href {http://arxiv.org/abs/#1} {{\tt arXiv:#1}}}
\def\mn@eprint@dblp#1{\href {http://dblp.uni-trier.de/rec/bibtex/#1.xml}
  {dblp:#1}}
\def\mn@eprint@#1:#2:#3:#4\@nil{\def\@tempa {#1}\def\@tempb {#2}\def\@tempc
  {#3}\ifx \@tempc \@empty \let \@tempc \@tempb \let \@tempb \@tempa \fi \ifx
  \@tempb \@empty \def\@tempb {arXiv}\fi \@ifundefined
  {mn@eprint@\@tempb}{\@tempb:\@tempc}{\expandafter \expandafter \csname
  mn@eprint@\@tempb\endcsname \expandafter{\@tempc}}}

\bibitem[\protect\citeauthoryear{{Altamirano}, {van der Klis}, {Wijnands}  \&
  {Cumming}}{{Altamirano} et~al.}{2008}]{art-2008-alta-all-a}
{Altamirano} D.,  {van der Klis} M.,  {Wijnands} R.,   {Cumming} A.,  2008,
  \mn@doi [\apjl] {10.1086/527355}, \href
  {https://ui.adsabs.harvard.edu/abs/2008ApJ...673L..35A} {673, L35}

\bibitem[\protect\citeauthoryear{{Belian}, {Conner}  \& {Evans}}{{Belian}
  et~al.}{1976}]{art-1976-beli-con-eva}
{Belian} R.~D.,  {Conner} J.~P.,   {Evans} W.~D.,  1976, \mn@doi [\apjl]
  {10.1086/182151}, \href {http://adsabs.harvard.edu/abs/1976ApJ...206L.135B}
  {206, L135}

\bibitem[\protect\citeauthoryear{{Bildsten}}{{Bildsten}}{1998}]{rev-1998-bild}
{Bildsten} L.,  1998, in {Buccheri} R.,  {van Paradijs} J.,   {Alpar} A.,  eds,
   NATO Advanced Science Institutes (ASI) Series C Vol. 515, NATO ASIC Proc.
  515: The Many Faces of Neutron Stars.. pp 419--+ (\mn@eprint {arXiv}
  {arXiv:astro-ph/9709094})

\bibitem[\protect\citeauthoryear{{Brown}}{{Brown}}{2004}]{art-2004-brow}
{Brown} E.~F.,  2004, \mn@doi [\apjl] {10.1086/425562}, \href
  {http://adsabs.harvard.edu/abs/2004ApJ...614L..57B} {614, L57}

\bibitem[\protect\citeauthoryear{{Brown} \& {Cumming}}{{Brown} \&
  {Cumming}}{2009}]{art-2009-brow-cum}
{Brown} E.~F.,  {Cumming} A.,  2009, \mn@doi [\apj]
  {10.1088/0004-637X/698/2/1020}, \href
  {http://adsabs.harvard.edu/abs/2009ApJ...698.1020B} {698, 1020}

\bibitem[\protect\citeauthoryear{{Cavecchi} \& {Spitkovsky}}{{Cavecchi} \&
  {Spitkovsky}}{2019}]{art-2019-cavecchi-spit}
{Cavecchi} Y.,  {Spitkovsky} A.,  2019, \mn@doi [\apj]
  {10.3847/1538-4357/ab3650}, \href
  {https://ui.adsabs.harvard.edu/abs/2019ApJ...882..142C} {882, 142}

\bibitem[\protect\citeauthoryear{{Cavecchi}, {Watts}, {Levin}  \&
  {Braithwaite}}{{Cavecchi} et~al.}{2015}]{art-2015-cavecchi-etal}
{Cavecchi} Y.,  {Watts} A.~L.,  {Levin} Y.,   {Braithwaite} J.,  2015, \mn@doi
  [\mnras] {10.1093/mnras/stu2764}, \href
  {http://adsabs.harvard.edu/abs/2015MNRAS.448..445C} {448, 445}

\bibitem[\protect\citeauthoryear{{Cavecchi}, {Watts}  \& {Galloway}}{{Cavecchi}
  et~al.}{2017}]{art-2017-cavecchi-etal}
{Cavecchi} Y.,  {Watts} A.~L.,   {Galloway} D.~K.,  2017, \mn@doi [\apj]
  {10.3847/1538-4357/aa9897}, \href
  {http://adsabs.harvard.edu/abs/2017ApJ...851....1C} {851, 1}

\bibitem[\protect\citeauthoryear{{Chevalier} \& {Ilovaisky}}{{Chevalier} \&
  {Ilovaisky}}{1991}]{art-1991-cheva-ilo}
{Chevalier} C.,  {Ilovaisky} S.~A.,  1991, \aap, \href
  {https://ui.adsabs.harvard.edu/abs/1991A&A...251L..11C} {251, L11}

\bibitem[\protect\citeauthoryear{{Cooper} \& {Narayan}}{{Cooper} \&
  {Narayan}}{2005}]{art-2005-coop-nara}
{Cooper} R.~L.,  {Narayan} R.,  2005, \mn@doi [\apj] {10.1086/431273}, \href
  {http://adsabs.harvard.edu/abs/2005ApJ...629..422C} {629, 422}

\bibitem[\protect\citeauthoryear{{Cooper} \& {Narayan}}{{Cooper} \&
  {Narayan}}{2007a}]{art-2007-coop-nara-a}
{Cooper} R.~L.,  {Narayan} R.,  2007a, \mn@doi [\apjl] {10.1086/513077}, \href
  {http://adsabs.harvard.edu/abs/2007ApJ...657L..29C} {657, L29}

\bibitem[\protect\citeauthoryear{{Cooper} \& {Narayan}}{{Cooper} \&
  {Narayan}}{2007b}]{art-2007-coop-nara-b}
{Cooper} R.~L.,  {Narayan} R.,  2007b, \mn@doi [\apj] {10.1086/513461}, \href
  {http://adsabs.harvard.edu/abs/2007ApJ...661..468C} {661, 468}

\bibitem[\protect\citeauthoryear{{Cornelisse} et~al.,}{{Cornelisse}
  et~al.}{2003}]{art-2003-corne-etal}
{Cornelisse} R.,  et~al., 2003, \mn@doi [\aap] {10.1051/0004-6361:20030629},
  \href {http://adsabs.harvard.edu/abs/2003A\%26A...405.1033C} {405, 1033}

\bibitem[\protect\citeauthoryear{{Crampton}, {Cowley}, {Stauffer}, {Ianna}  \&
  {Hutchings}}{{Crampton} et~al.}{1986}]{art-1986-cramp-etal}
{Crampton} D.,  {Cowley} A.~P.,  {Stauffer} J.,  {Ianna} P.,   {Hutchings}
  J.~B.,  1986, \mn@doi [\apj] {10.1086/164369}, \href
  {https://ui.adsabs.harvard.edu/abs/1986ApJ...306..599C} {306, 599}

\bibitem[\protect\citeauthoryear{{Cumming}}{{Cumming}}{2003}]{art-2003-cum}
{Cumming} A.,  2003, \mn@doi [\apj] {10.1086/377446}, \href
  {http://adsabs.harvard.edu/abs/2003ApJ...595.1077C} {595, 1077}

\bibitem[\protect\citeauthoryear{{Cumming}, {Macbeth}, {in't Zand}  \&
  {Page}}{{Cumming} et~al.}{2006}]{art-2006-cum-mac-tzan-page}
{Cumming} A.,  {Macbeth} J.,  {in't Zand} J.~J.~M.,   {Page} D.,  2006, \mn@doi
  [\apj] {10.1086/504698}, \href
  {http://adsabs.harvard.edu/abs/2006ApJ...646..429C} {646, 429}

\bibitem[\protect\citeauthoryear{{Cyburt}, {Amthor}, {Heger}, {Johnson},
  {Keek}, {Meisel}, {Schatz}  \& {Smith}}{{Cyburt}
  et~al.}{2016}]{art-2016-cyb-etal}
{Cyburt} R.~H.,  {Amthor} A.~M.,  {Heger} A.,  {Johnson} E.,  {Keek} L.,
  {Meisel} Z.,  {Schatz} H.,   {Smith} K.,  2016, \mn@doi [\apj]
  {10.3847/0004-637X/830/2/55}, \href
  {http://adsabs.harvard.edu/abs/2016ApJ...830...55C} {830, 55}

\bibitem[\protect\citeauthoryear{{Galloway}, {Muno}, {Hartman}, {Psaltis}  \&
  {Chakrabarty}}{{Galloway} et~al.}{2008}]{rev-2008-gal-mun-hart-psal-chak}
{Galloway} D.~K.,  {Muno} M.~P.,  {Hartman} J.~M.,  {Psaltis} D.,
  {Chakrabarty} D.,  2008, \mn@doi [\apjs] {10.1086/592044}, \href
  {http://adsabs.harvard.edu/abs/2008ApJS..179..360G} {179, 360}

\bibitem[\protect\citeauthoryear{{Galloway}, {Chakrabarty}  \&
  {Lin}}{{Galloway} et~al.}{2009}]{atel-2009-gal-etal}
{Galloway} D.~K.,  {Chakrabarty} D.,   {Lin} J.~R.,  2009, The Astronomer's
  Telegram, \href {https://ui.adsabs.harvard.edu/abs/2009ATel.2094....1G}
  {2094, 1}

\bibitem[\protect\citeauthoryear{{Galloway}, {Yao}, {Marshall}, {Misanovic}  \&
  {Weinberg}}{{Galloway} et~al.}{2010}]{art-2010-gall-etal}
{Galloway} D.~K.,  {Yao} Y.,  {Marshall} H.,  {Misanovic} Z.,   {Weinberg} N.,
  2010, \mn@doi [\apj] {10.1088/0004-637X/724/1/417}, \href
  {https://ui.adsabs.harvard.edu/abs/2010ApJ...724..417G} {724, 417}

\bibitem[\protect\citeauthoryear{{Galloway} et~al.,}{{Galloway}
  et~al.}{2018}]{art-2018-gal-etal}
{Galloway} D.~K.,  et~al., 2018, \mn@doi [\apjl] {10.3847/2041-8213/aabd32},
  \href {http://adsabs.harvard.edu/abs/2018ApJ...857L..24G} {857, L24}

\bibitem[\protect\citeauthoryear{{Galloway} et~al.,}{{Galloway}
  et~al.}{2020}]{rev-arx-2020-gall-etal}
{Galloway} D.~K.,  et~al., 2020, arXiv e-prints, \href
  {https://ui.adsabs.harvard.edu/abs/2020arXiv200300685G} {p. arXiv:2003.00685}

\bibitem[\protect\citeauthoryear{{Goodwin}, {Galloway}, {Heger}, {Cumming}  \&
  {Johnston}}{{Goodwin} et~al.}{2019a}]{art-2019-goodw-etal}
{Goodwin} A.~J.,  {Galloway} D.~K.,  {Heger} A.,  {Cumming} A.,   {Johnston}
  Z.,  2019a, \mn@doi [\mnras] {10.1093/mnras/stz2638}, \href
  {https://ui.adsabs.harvard.edu/abs/2019MNRAS.490.2228G} {490, 2228}

\bibitem[\protect\citeauthoryear{{Goodwin}, {Heger}  \& {Galloway}}{{Goodwin}
  et~al.}{2019b}]{art-2019-goodw-heg-gal}
{Goodwin} A.~J.,  {Heger} A.,   {Galloway} D.~K.,  2019b, \mn@doi [\apj]
  {10.3847/1538-4357/aaeed2}, \href
  {http://adsabs.harvard.edu/abs/2019ApJ...870...64G} {870, 64}

\bibitem[\protect\citeauthoryear{{Grebenev} \& {Chelovekov}}{{Grebenev} \&
  {Chelovekov}}{2018}]{art-arx-2018-greb-chelo}
{Grebenev} S.~A.,  {Chelovekov} I.~V.,  2018, \mn@doi [Astronomy Letters]
  {10.1134/S1063773718120083}, \href
  {https://ui.adsabs.harvard.edu/abs/2018AstL...44..777G} {44, 777}

\bibitem[\protect\citeauthoryear{{He} \& {Keek}}{{He} \& {Keek}}{2016}]{he16}
{He} C.-C.,  {Keek} L.,  2016, \mn@doi [\apj] {10.3847/0004-637X/819/1/47},
  \href {http://adsabs.harvard.edu/abs/2016ApJ...819...47H} {819, 47}

\bibitem[\protect\citeauthoryear{{Heger}, {Cumming}  \& {Woosley}}{{Heger}
  et~al.}{2007a}]{art-2007-heg-cum-woos}
{Heger} A.,  {Cumming} A.,   {Woosley} S.~E.,  2007a, \mn@doi [\apj]
  {10.1086/517491}, \href {http://adsabs.harvard.edu/abs/2007ApJ...665.1311H}
  {665, 1311}

\bibitem[\protect\citeauthoryear{{Heger}, {Cumming}, {Galloway}  \&
  {Woosley}}{{Heger} et~al.}{2007b}]{art-2007-heg-cumm-gal-woos}
{Heger} A.,  {Cumming} A.,  {Galloway} D.~K.,   {Woosley} S.~E.,  2007b,
  \mn@doi [\apjl] {10.1086/525522}, \href
  {http://adsabs.harvard.edu/abs/2007ApJ...671L.141H} {671, L141}

\bibitem[\protect\citeauthoryear{{Hoffman}, {Lewin}, {Doty}, {Hearn}, {Clark},
  {Jernigan}  \& {Li}}{{Hoffman} et~al.}{1976}]{art-1976-hoff-lew-etal}
{Hoffman} J.~A.,  {Lewin} W.~H.~G.,  {Doty} J.,  {Hearn} D.~R.,  {Clark} G.~W.,
   {Jernigan} G.,   {Li} F.~K.,  1976, \mn@doi [\apjl] {10.1086/182292}, \href
  {https://ui.adsabs.harvard.edu/abs/1976ApJ...210L..13H} {210, L13}

\bibitem[\protect\citeauthoryear{{Inogamov} \& {Sunyaev}}{{Inogamov} \&
  {Sunyaev}}{1999}]{art-1999-ino-suny}
{Inogamov} N.~A.,  {Sunyaev} R.~A.,  1999, Astronomy Letters, \href
  {http://adsabs.harvard.edu/abs/1999AstL...25..269I} {25, 269}

\bibitem[\protect\citeauthoryear{{Johnston}, {Heger}  \& {Galloway}}{{Johnston}
  et~al.}{2018}]{art-2018-johnst-etal}
{Johnston} Z.,  {Heger} A.,   {Galloway} D.~K.,  2018, \mn@doi [\mnras]
  {10.1093/mnras/sty757}, \href
  {https://ui.adsabs.harvard.edu/abs/2018MNRAS.477.2112J} {477, 2112}

\bibitem[\protect\citeauthoryear{{Kajava} et~al.,}{{Kajava}
  et~al.}{2014}]{art-2014-kaja-etal}
{Kajava} J.~J.~E.,  et~al., 2014, \mn@doi [\mnras] {10.1093/mnras/stu2073},
  \href {http://adsabs.harvard.edu/abs/2014MNRAS.445.4218K} {445, 4218}

\bibitem[\protect\citeauthoryear{{Keek} \& {Heger}}{{Keek} \&
  {Heger}}{2011}]{art-2011-keek-heger}
{Keek} L.,  {Heger} A.,  2011, \mn@doi [\apj] {10.1088/0004-637X/743/2/189},
  \href {http://adsabs.harvard.edu/abs/2011ApJ...743..189K} {743, 189}

\bibitem[\protect\citeauthoryear{{Keek} \& {Heger}}{{Keek} \&
  {Heger}}{2016}]{art-2016-kee-heg}
{Keek} L.,  {Heger} A.,  2016, \mn@doi [\mnras] {10.1093/mnrasl/slv167}, \href
  {http://adsabs.harvard.edu/abs/2016MNRAS.456L..11K} {456, L11}

\bibitem[\protect\citeauthoryear{{Keek} \& {Heger}}{{Keek} \&
  {Heger}}{2017}]{KH17}
{Keek} L.,  {Heger} A.,  2017, \mn@doi [\apj] {10.3847/1538-4357/aa7748}, \href
  {https://ui.adsabs.harvard.edu/abs/2017ApJ...842..113K} {842, 113}

\bibitem[\protect\citeauthoryear{{Keek}, {Langer}  \& {in't Zand}}{{Keek}
  et~al.}{2009}]{art-2009-kee-lang-zand}
{Keek} L.,  {Langer} N.,   {in't Zand} J.~J.~M.,  2009, \mn@doi [\aap]
  {10.1051/0004-6361/200911619}, \href
  {http://adsabs.harvard.edu/abs/2009A%26A...502..871K} {502, 871}

\bibitem[\protect\citeauthoryear{{Keek}, {Galloway}, {in't Zand}  \&
  {Heger}}{{Keek} et~al.}{2010}]{art-2010-keek-etal}
{Keek} L.,  {Galloway} D.~K.,  {in't Zand} J.~J.~M.,   {Heger} A.,  2010,
  \mn@doi [\apj] {10.1088/0004-637X/718/1/292}, \href
  {http://adsabs.harvard.edu/abs/2010ApJ...718..292K} {718, 292}

\bibitem[\protect\citeauthoryear{{Keek}, {Cyburt}  \& {Heger}}{{Keek}
  et~al.}{2014}]{art-2014-keek-cyb-heger}
{Keek} L.,  {Cyburt} R.~H.,   {Heger} A.,  2014, \mn@doi [\apj]
  {10.1088/0004-637X/787/2/101}, \href
  {http://adsabs.harvard.edu/abs/2014ApJ...787..101K} {787, 101}

\bibitem[\protect\citeauthoryear{{Lampe}, {Heger}  \& {Galloway}}{{Lampe}
  et~al.}{2016}]{art-2016-lampe-heg-gal}
{Lampe} N.,  {Heger} A.,   {Galloway} D.~K.,  2016, \mn@doi [\apj]
  {10.3847/0004-637X/819/1/46}, \href
  {http://adsabs.harvard.edu/abs/2016ApJ...819...46L} {819, 46}

\bibitem[\protect\citeauthoryear{{Lewin}, {Li}, {Hoffman}, {Doty}, {Buff},
  {Clark}  \& {Rappaport}}{{Lewin} et~al.}{1976}]{art-1976-lew-etal}
{Lewin} W.~H.~G.,  {Li} F.~K.,  {Hoffman} J.~A.,  {Doty} J.,  {Buff} J.,
  {Clark} G.~W.,   {Rappaport} S.,  1976, \mn@doi [\mnras]
  {10.1093/mnras/177.1.93P}, \href
  {https://ui.adsabs.harvard.edu/abs/1976MNRAS.177P..93L} {177, 93P}

\bibitem[\protect\citeauthoryear{{Lewin}, {van Paradijs}  \& {Taam}}{{Lewin}
  et~al.}{1993}]{rev-1993-lew-par-taa}
{Lewin} W.~H.~G.,  {van Paradijs} J.,   {Taam} R.~E.,  1993, \mn@doi [Space
  Science Reviews] {10.1007/BF00196124}, \href
  {http://adsabs.harvard.edu/abs/1993SSRv...62..223L} {62, 223}

\bibitem[\protect\citeauthoryear{{Linares}, {Altamirano}, {Chakrabarty},
  {Cumming}  \& {Keek}}{{Linares} et~al.}{2012}]{art-2012-lin-etal}
{Linares} M.,  {Altamirano} D.,  {Chakrabarty} D.,  {Cumming} A.,   {Keek} L.,
  2012, \mn@doi [\apj] {10.1088/0004-637X/748/2/82}, \href
  {http://adsabs.harvard.edu/abs/2012ApJ...748...82L} {748, 82}

\bibitem[\protect\citeauthoryear{{Lyu}, {M{\'e}ndez}, {Zhang}  \& {Keek}}{{Lyu}
  et~al.}{2015}]{art-2015-lyu-etal}
{Lyu} M.,  {M{\'e}ndez} M.,  {Zhang} G.,   {Keek} L.,  2015, \mn@doi [\mnras]
  {10.1093/mnras/stv1971}, \href
  {http://adsabs.harvard.edu/abs/2015MNRAS.454..541L} {454, 541}

\bibitem[\protect\citeauthoryear{{Markwardt}, {Strohmayer}  \&
  {Swank}}{{Markwardt} et~al.}{1999}]{art-1999-mark-stroh-etal}
{Markwardt} C.~B.,  {Strohmayer} T.~E.,   {Swank} J.~H.,  1999, \mn@doi [\apjl]
  {10.1086/311886}, \href
  {https://ui.adsabs.harvard.edu/abs/1999ApJ...512L.125M} {512, L125}

\bibitem[\protect\citeauthoryear{{Marshall}, {Li}  \& {Rappaport}}{{Marshall}
  et~al.}{1977}]{art-1977-mars-li-rappa}
{Marshall} H.,  {Li} F.,   {Rappaport} S.,  1977, \iaucirc, \href
  {https://ui.adsabs.harvard.edu/abs/1977IAUC.3134....2M} {3134, 2}

\bibitem[\protect\citeauthoryear{{Meisel}, {Deibel}, {Keek}, {Shternin}  \&
  {Elfritz}}{{Meisel} et~al.}{2018}]{rev-2018-meis-etal}
{Meisel} Z.,  {Deibel} A.,  {Keek} L.,  {Shternin} P.,   {Elfritz} J.,  2018,
  \mn@doi [Journal of Physics G Nuclear Physics] {10.1088/1361-6471/aad171},
  \href {http://adsabs.harvard.edu/abs/2018JPhG...45i3001M} {45, 093001}

\bibitem[\protect\citeauthoryear{{Misanovic}, {Galloway}  \&
  {Cooper}}{{Misanovic} et~al.}{2010}]{art-2010-misa-gall-coop}
{Misanovic} Z.,  {Galloway} D.~K.,   {Cooper} R.~L.,  2010, \mn@doi [\apj]
  {10.1088/0004-637X/718/2/947}, \href
  {https://ui.adsabs.harvard.edu/abs/2010ApJ...718..947M} {718, 947}

\bibitem[\protect\citeauthoryear{{Muno}, {Fox}, {Morgan}  \& {Bildsten}}{{Muno}
  et~al.}{2000}]{art-2000-muno-etal}
{Muno} M.~P.,  {Fox} D.~W.,  {Morgan} E.~H.,   {Bildsten} L.,  2000, \mn@doi
  [\apj] {10.1086/317031}, \href
  {https://ui.adsabs.harvard.edu/abs/2000ApJ...542.1016M} {542, 1016}

\bibitem[\protect\citeauthoryear{{Muno}, {Chakrabarty}, {Galloway}  \&
  {Psaltis}}{{Muno} et~al.}{2002}]{art-2002-muno-etal}
{Muno} M.~P.,  {Chakrabarty} D.,  {Galloway} D.~K.,   {Psaltis} D.,  2002,
  \mn@doi [\apj] {10.1086/343793}, \href
  {http://adsabs.harvard.edu/abs/2002ApJ...580.1048M} {580, 1048}

\bibitem[\protect\citeauthoryear{{Narayan} \& {Heyl}}{{Narayan} \&
  {Heyl}}{2003}]{art-2003-nara-heyl}
{Narayan} R.,  {Heyl} J.~S.,  2003, \mn@doi [\apj] {10.1086/379211}, \href
  {http://adsabs.harvard.edu/abs/2003ApJ...599..419N} {599, 419}

\bibitem[\protect\citeauthoryear{{Paczynski}}{{Paczynski}}{1983}]{Pac83}
{Paczynski} B.,  1983, \mn@doi [\apj] {10.1086/160596}, \href
  {https://ui.adsabs.harvard.edu/abs/1983ApJ...264..282P} {264, 282}

\bibitem[\protect\citeauthoryear{{Parmar}, {Gottwald}, {Haberl}  \&
  {White}}{{Parmar} et~al.}{1985}]{atel-1985-par-etal}
{Parmar} A.~N.,  {Gottwald} M.,  {Haberl} F.,   {White} N.~E.,  1985, \iaucirc,
  \href {https://ui.adsabs.harvard.edu/abs/1985IAUC.4057....1P} {4057}

\bibitem[\protect\citeauthoryear{{Piro} \& {Bildsten}}{{Piro} \&
  {Bildsten}}{2007}]{art-2007-piro-bild}
{Piro} A.~L.,  {Bildsten} L.,  2007, \mn@doi [\apj] {10.1086/518687}, \href
  {http://adsabs.harvard.edu/abs/2007ApJ...663.1252P} {663, 1252}

\bibitem[\protect\citeauthoryear{{Ray} et~al.,}{{Ray}
  et~al.}{2019}]{art-arx-2019-ray-etal}
{Ray} P.~S.,  et~al., 2019, arXiv e-prints, \href
  {http://adsabs.harvard.edu/abs/2019arXiv190303035R} {}

\bibitem[\protect\citeauthoryear{{Revnivtsev}, {Churazov}, {Gilfanov}  \&
  {Sunyaev}}{{Revnivtsev} et~al.}{2001}]{art-2001-rev-etal}
{Revnivtsev} M.,  {Churazov} E.,  {Gilfanov} M.,   {Sunyaev} R.,  2001, \mn@doi
  [\aap] {10.1051/0004-6361:20010434}, \href
  {https://ui.adsabs.harvard.edu/abs/2001A&A...372..138R} {372, 138}

\bibitem[\protect\citeauthoryear{{Smith}, {Morgan}  \& {Bradt}}{{Smith}
  et~al.}{1997}]{art-1997-smi-etal}
{Smith} D.~A.,  {Morgan} E.~H.,   {Bradt} H.,  1997, \mn@doi [\apjl]
  {10.1086/310604}, \href
  {https://ui.adsabs.harvard.edu/abs/1997ApJ...479L.137S} {479, L137}

\bibitem[\protect\citeauthoryear{{Strohmayer}, {Zhang}, {Swank}, {Smale},
  {Titarchuk}, {Day}  \& {Lee}}{{Strohmayer} et~al.}{1996}]{art-1996-stro-etal}
{Strohmayer} T.~E.,  {Zhang} W.,  {Swank} J.~H.,  {Smale} A.,  {Titarchuk} L.,
  {Day} C.,   {Lee} U.,  1996, \mn@doi [\apjl] {10.1086/310261}, \href
  {http://adsabs.harvard.edu/abs/1996ApJ...469L...9S} {469, L9+}

\bibitem[\protect\citeauthoryear{{Strohmayer}, {Zhang}, {Swank}, {White}  \&
  {Lapidus}}{{Strohmayer} et~al.}{1998}]{art-1998-stroh-etal}
{Strohmayer} T.~E.,  {Zhang} W.,  {Swank} J.~H.,  {White} N.~E.,   {Lapidus}
  I.,  1998, \mn@doi [\apjl] {10.1086/311322}, \href
  {https://ui.adsabs.harvard.edu/abs/1998ApJ...498L.135S} {498, L135}

\bibitem[\protect\citeauthoryear{{Sunyaev} \& {Kwant Team}}{{Sunyaev} \& {Kwant
  Team}}{1989}]{atel-1989-suny-kwant}
{Sunyaev} R.,  {Kwant Team} 1989, \iaucirc, \href
  {https://ui.adsabs.harvard.edu/abs/1989IAUC.4839....1S} {4839}

\bibitem[\protect\citeauthoryear{{Swank}, {Becker}, {Pravdo}, {Saba}  \&
  {Serlemitsos}}{{Swank} et~al.}{1976}]{atel-1976-swank-etal}
{Swank} J.~H.,  {Becker} R.~H.,  {Pravdo} S.~H.,  {Saba} J.~R.,   {Serlemitsos}
  P.~J.,  1976, \iaucirc, \href
  {https://ui.adsabs.harvard.edu/abs/1976IAUC.3000U...1S} {3000}

\bibitem[\protect\citeauthoryear{{Vincentelli}, {Cavecchi}, {Casella},
  {Migliari}, {Altamirano}, {Belloni}  \& {Diaz-Trigo}}{{Vincentelli}
  et~al.}{2020}]{art-2020-vinc-cavecchi-etal}
{Vincentelli} F.~M.,  {Cavecchi} Y.,  {Casella} P.,  {Migliari} S.,
  {Altamirano} D.,  {Belloni} T.,   {Diaz-Trigo} M.,  2020, \mn@doi [\mnras]
  {10.1093/mnrasl/slaa049}, \href
  {https://ui.adsabs.harvard.edu/abs/2020MNRAS.495L..37V} {495, L37}

\bibitem[\protect\citeauthoryear{{Wachter}, {Hoard}, {Bailyn}, {Corbel}  \&
  {Kaaret}}{{Wachter} et~al.}{2002}]{art-2002-wach-etal}
{Wachter} S.,  {Hoard} D.~W.,  {Bailyn} C.~D.,  {Corbel} S.,   {Kaaret} P.,
  2002, \mn@doi [\apj] {10.1086/339034}, \href
  {https://ui.adsabs.harvard.edu/abs/2002ApJ...568..901W} {568, 901}

\bibitem[\protect\citeauthoryear{{Wijnands}, {Degenaar}  \& {Page}}{{Wijnands}
  et~al.}{2017}]{rev-2017-wij-dege-page}
{Wijnands} R.,  {Degenaar} N.,   {Page} D.,  2017, \mn@doi [Journal of
  Astrophysics and Astronomy] {10.1007/s12036-017-9466-5}, \href
  {http://adsabs.harvard.edu/abs/2017JApA...38...49W} {38, 49}

\bibitem[\protect\citeauthoryear{{Woosley} et~al.,}{{Woosley}
  et~al.}{2004}]{art-2004-woos-etal}
{Woosley} S.~E.,  et~al., 2004, \mn@doi [\apjs] {10.1086/381533}, \href
  {http://adsabs.harvard.edu/abs/2004ApJS..151...75W} {151, 75}

\bibitem[\protect\citeauthoryear{{Zamfir}, {Cumming}  \& {Niquette}}{{Zamfir}
  et~al.}{2014}]{Zam14}
{Zamfir} M.,  {Cumming} A.,   {Niquette} C.,  2014, \mn@doi [\mnras]
  {10.1093/mnras/stu1927}, \href
  {https://ui.adsabs.harvard.edu/abs/2014MNRAS.445.3278Z} {445, 3278}

\bibitem[\protect\citeauthoryear{{Zhang}, {Jahoda}, {Kelley}, {Strohmayer},
  {Swank}  \& {Zhang}}{{Zhang} et~al.}{1998}]{art-1998-zhang-all}
{Zhang} W.,  {Jahoda} K.,  {Kelley} R.~L.,  {Strohmayer} T.~E.,  {Swank} J.~H.,
    {Zhang} S.~N.,  1998, \mn@doi [\apjl] {10.1086/311210}, \href
  {http://adsabs.harvard.edu/abs/1998ApJ...495L...9Z} {495, L9+}

\bibitem[\protect\citeauthoryear{{Zhang} et~al.,}{{Zhang}
  et~al.}{2019}]{art-2018-zhang-etal}
{Zhang} S.,  et~al., 2019, \mn@doi [Science China Physics, Mechanics, and
  Astronomy] {10.1007/s11433-018-9309-2}, \href
  {https://ui.adsabs.harvard.edu/abs/2019SCPMA..6229502Z} {62, 29502}

\bibitem[\protect\citeauthoryear{{in't Zand}, {Jonker}  \& {Markwardt}}{{in't
  Zand} et~al.}{2007}]{art-2007-zand-jon-markw}
{in't Zand} J.~J.~M.,  {Jonker} P.~G.,   {Markwardt} C.~B.,  2007, \mn@doi
  [\aap] {10.1051/0004-6361:20066678}, \href
  {http://adsabs.harvard.edu/abs/2007A\%26A...465..953I} {465, 953}

\bibitem[\protect\citeauthoryear{{van Paradijs}, {Penninx}  \& {Lewin}}{{van
  Paradijs} et~al.}{1988}]{art-1988-par-pen-lew}
{van Paradijs} J.,  {Penninx} W.,   {Lewin} W.~H.~G.,  1988, \mn@doi [\mnras]
  {10.1093/mnras/233.2.437}, \href
  {http://adsabs.harvard.edu/abs/1988MNRAS.233..437V} {233, 437}

\bibitem[\protect\citeauthoryear{{van Paradijs} et~al.,}{{van Paradijs}
  et~al.}{1990}]{art-1990-par-etal}
{van Paradijs} J.,  et~al., 1990, \aap, \href
  {https://ui.adsabs.harvard.edu/abs/1990A&A...234..181V} {234, 181}

\makeatother
\end{thebibliography}

\label{lastpage}

\end{document}